\RequirePackage[T1]{fontenc}
\documentclass[10pt,journal,letterpaper]{IEEEtran}
\usepackage{amsmath,amsfonts,amssymb}
\usepackage{algpseudocode,algorithm}
\algrenewcommand\algorithmicrequire{\textbf{Input:}}
\algrenewcommand\algorithmicensure{\textbf{Output:}}
\usepackage[caption=false,font=footnotesize]{subfig}
\usepackage{textcomp,stfloats,url,graphicx,cite,booktabs,tabularx,array,placeins}
\usepackage[hidelinks,implicit=false,bookmarks=false]{hyperref}
\graphicspath{{figures/}}
\newtheorem{proposition}{Proposition}
\AtBeginDocument{\setlength{\abovedisplayskip}{5pt plus 1.5pt minus 2pt}\setlength{\belowdisplayskip}{5pt plus 1.5pt minus 2pt}\setlength{\abovedisplayshortskip}{2pt plus 1pt}\setlength{\belowdisplayshortskip}{2.5pt plus 1pt}}

\begin{document}
\title{Motion-Native Predictive Beam Shaping via Geometric Entities for Near-Field XL-MIMO}
\author{Yuheng Fan, Chuang Yang,~\IEEEmembership{Member,~IEEE,}
and Mugen Peng,~\IEEEmembership{Fellow,~IEEE}%
\thanks{This work was supported by the National Natural Science Foundation of China under Grant No. 62671067, in part by the BUPT Excellent Ph.D. Students Foundation under Grant CX20251017, in part by the BUPT Innovation and Entrepreneurship Support Program under Grant 2025-YC-T001, and in part by the Doctoral Student Program of the Young S$\&$T Talents Cultivation Project, CAST. \textit{(Corresponding author: Chuang Yang.)}

Y. Fan and M. Peng are with the State Key Laboratory of Networking and Switching Technology, Beijing University of Posts and Telecommunications, Beijing 100876, China (e-mail: vanyh@bupt.edu.cn; pmg@bupt.edu.cn).

C. Yang is with the State Key Laboratory of Networking and Switching Technology, Beijing University of Posts and Telecommunications, Beijing 100876, China, and also with the Xiong'an Aerospace Information Research Institute, Xiong'an New Area 071700, China (e-mail: chuangyang@\allowbreak bupt.edu.cn).}}
\markboth{}{}
\AddToHookNext{shipout/foreground}{%
  \put(0,\dimexpr-\paperheight+19bp\relax){%
    \makebox[\paperwidth][c]{\parbox[b]{\textwidth}{\centering\normalfont\footnotesize
      This work will be submitted to the IEEE for possible publication.\\
      Copyright may be transferred without notice, after which this version may no longer be accessible.}}}%
}
\maketitle

\begin{abstract}
In extremely large-scale multiple-input multiple-output (XL-MIMO) systems, spherical-wave propagation enables tight range-angle focusing, yet finite updates, execution delays, and cross-interval weight transitions render near-field links vulnerable to user motion. Point-focused beams rapidly lose alignment across a service interval, whereas unaligned updates induce handoff notches. We propose geometrically parameterized beamshaping (GPBS) to bridge asynchronous heterogeneous observations and cross-window spatial control. Specifically, a continuous-time geometric entity fuses complex pilots with delayed point and interval sensing on their native temporal supports, guides single-pass pilot likelihood refinement, and supplies terminal derivatives for quadratic forecasting. GPBS then coherently integrates predicted spherical responses under a cross-window activation kernel with constructive handoff phase alignment in $O(MQ)$ operations. Theoretically, we characterize the spatial directions resolved by gain-profiled pilots, establish when fusing pilots with interval reports lowers modeled forecast variance below both single-source fits, and derive the exact condition for normalized beam transitions to exceed both constituent gains. Simulations demonstrate substantial improvements in spectral efficiency and rate continuity over causal tracking and alternative beam-shaping baselines, achieving millisecond-level beam synthesis with consistent gains across diverse mobility profiles, service-timing regimes, and near-field multipath channels.
\end{abstract}
\begin{IEEEkeywords}
Geometric entities, near-field XL-MIMO, predictive beam shaping, motion-aware communications, predictive beam control, heterogeneous sensing.
\end{IEEEkeywords}

\section{Introduction}
\label{sec:introduction}
\IEEEPARstart{E}{xtremely} large-scale multiple-input multiple-output (XL-MIMO) base stations expand physical array apertures to support high-rate wireless communications \cite{Wang2024XLMIMO,Lu2024Tutorial,Cui2023NearField}. Larger apertures increase Rayleigh distances, making radiative near-field propagation relevant over a wider range of link distances. In this region, spherical wavefront curvature makes the relative propagation phases across the array depend on both the user's range and angle. By adjusting the transmit weights to compensate for these phase differences, the base station can make signals from different elements add coherently at a desired location, producing a spatial focus. However, this enhanced focusing can make the received gain more sensitive to user motion when the beam remains focused on the user's earlier position. 

\begin{figure*}[!t]
\centering
\includegraphics[width=\textwidth]{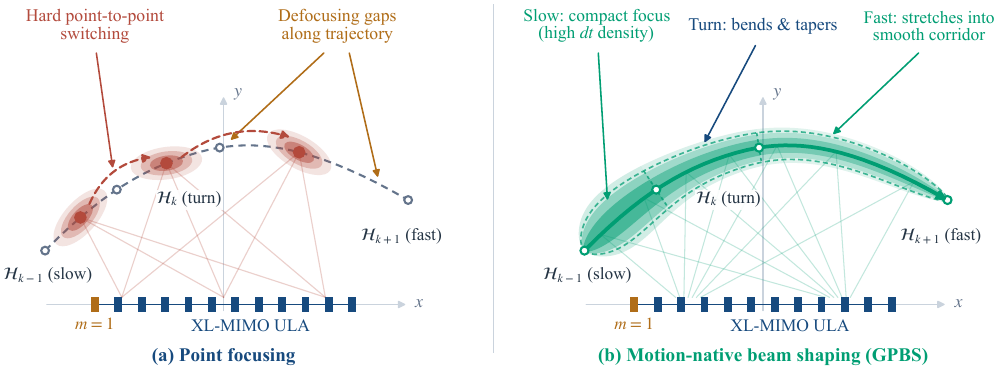}
\caption{Schematic comparison of point focusing and GPBS across service intervals.}
\label{fig:fig1}
\end{figure*}

Maintaining this focus under mobility is constrained by asynchronous pilot and sensing schedules, beam computation~\cite{Nazar2026Enwar3}, and hardware reconfiguration latencies from antenna movement, liquid-crystal relaxation, or port switching~\cite{Wang2026MAMovement,Wang2024LiquidCrystal,Dinis2026FASSwitching}. Consequently, a beam selected at a decision time is activated after an execution delay and held fixed over a finite service interval. Because the user continues moving across both the execution delay and the hold, aiming each beam at a single predicted position leaves much of the interval unfocused, as shown in Fig.~\ref{fig:fig1}(a). This motivates \emph{motion-native beam shaping}, illustrated in Fig.~\ref{fig:fig1}(b), in which predicted motion geometry and service timing jointly shape one fixed beam per interval.

\subsection{Related Work and Research Motivation}
Empirical studies of operational beam management, positioning, on-device inference, and reconfigurable apertures confirm non-negligible reporting, computation, and switching latencies~\cite{Feng2025BeamManagement,Hernandez2024GNSSLatency,Urbinati2026RISLatency,Biswas2026VIBE,Wang2026MAMovement,Dinis2026FASSwitching,Ting2026LCRIS}, over which near-field spherical channels decorrelate rapidly under user motion~\cite{Zhang2026NearFieldDynamics}. Existing near-field predictive schemes estimate radial and transverse velocities from echoes, track kinematic states across sensing slots, or adapt beam selection via payload feedback~\cite{Wang2025Velocity,Jiang2024NISE,Liu2026Thompson}. However, these designs still focus each beam on a single predicted location rather than shaping coverage across a held service interval.

To improve coverage over an update interval, Chen et al.~\cite{Chen2024Adaptive} coherently integrate far-field steering vectors over a predicted angular sweep, deriving $\operatorname{sinc}$-type element weights with an optimizable phase offset selected via an offline codebook. Their formulation assumes far-field plane waves and uniform angular sweeps, leaving spherical range-angle coupling and residence-time weighting unmodeled. In the near field, Ye et al.~\cite{Ye2025TABS} shape aperture phases along a prescribed spatial path by arc-length coverage, without modeling causal trajectory inference or physical-time occupancy.

Our earlier work~\cite{Fan2026GeometricTracking} introduced B-spline geometric entities and gain-profiled pilot likelihoods for instantaneous THz channel tracking without execution delay, interval-operator sensing, or held-beam synthesis. Building on continuous-time trajectory representations~\cite{Barfoot2014SparseGP,Johnson2024Continuous,deBoor2001}, this paper bridges asynchronous multi-rate inference and cross-window near-field beam shaping across delayed service intervals.

\subsection{Motion-Native Beam Shaping via Geometric Entities}
We develop geometrically parameterized beamshaping (GPBS) across delayed service intervals. The main contributions of this paper are summarized as follows.
\begin{enumerate}
\item We construct a continuous geometric entity in physical time to fuse asynchronous pilots and point and interval sensing on their native temporal supports. Entity-guided pilot refinement and terminal-derivative readout connect the fused motion history to an order-$N$ forecast over the delayed service interval.
\item We characterize when asynchronous pilots and interval reports provide complementary forecast information. Under the local Gaussian model with fixed pilot factors, their joint fit reduces directional forecast variance below both single-source fits exactly when each source contributes forecast-correlated information beyond the other.
\item We develop cross-window GPBS by coherently integrating the predicted spherical responses under a physical-time activation kernel and aligning adjacent beams at the predicted handoff position. The closed-form beam has $O(MQ)$ synthesis complexity. We characterize when its normalized transition yields higher gain than either constituent beam and evaluate the framework with paired system comparisons, ablations, and sensitivity studies.
\end{enumerate}

The remainder of this paper is organized as follows. Section~\ref{sec:system} formulates the system model and causal beam shaping task. Section~\ref{sec:inference} develops historical entity inference and pilot refinement, Section~\ref{sec:beam} derives terminal forecasting and closed-form GPBS, Section~\ref{sec:results} presents the simulation results, and Section~\ref{sec:conclusion} concludes the paper.

\section{System Model and Causal Beam Shaping Task}
\label{sec:system}
\begin{figure}[!t]
\centering
\includegraphics[width=0.80\columnwidth]{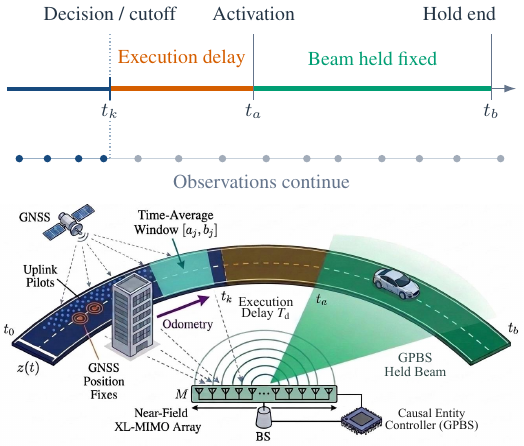}
\caption{System scenario and beam-control timing.}
\label{fig:fig2}
\end{figure}

As illustrated in Fig.~\ref{fig:fig2}, we consider a near-field downlink system where a base station (BS) with an $M$-element uniform linear array (ULA) serves mobile single-antenna users over orthogonal time slots and $L$ orthogonal subcarriers. Without loss of generality, we focus on one representative user and subcarrier, omitting both indices. A beam controller processes full-array uplink pilots alongside heterogeneous point and interval sensing reports to select the BS transmit weights for each service interval.

\subsection{Near-Field Channel and Beamforming Model}

The BS array lies along the $x$ axis centered at the origin with element coordinates $\mathbf x=[x_1,\ldots,x_M]^{\mathsf T}$, where
\begin{equation}
x_m=\left(m-\frac{M+1}{2}\right)d,\quad m=1,\ldots,M,
\label{eq:array_coordinates}
\end{equation}
with fixed spacing $d=\lambda/2$, wavelength $\lambda$, and aperture $D=(M-1)d$.

At time $t$, the user lies at polar coordinates $(\rho(t),\theta(t))$, with range $\rho(t)>0$ from the array center and angle $\theta(t)$ from the positive $x$ axis, giving Cartesian position
\begin{equation}
\mathbf z(t)=[\rho(t)\cos\theta(t),\rho(t)\sin\theta(t)]^{\mathsf T}
\label{eq:cartesian_position}
\end{equation}
and element propagation distances $\mathbf r=[r_1,\ldots,r_M]^{\mathsf T}$ with
\begin{equation}
r_m=\sqrt{\rho^2+x_m^2-2\rho x_m\cos\theta}.
\label{eq:element_range}
\end{equation}
Writing $r_m(t)$ along the motion, the normalized response of element $m$ in a narrowband single-path near-field channel with equal element amplitudes~\cite{Lu2024Tutorial,Cui2023NearField} depends on its propagation distance relative to that of element $1$ as
\begin{equation}
a_m(\rho,\theta)
=\frac{1}{\sqrt M}\exp\!\left[
-j\frac{2\pi}{\lambda}\bigl(r_m-r_1\bigr)
\right].
\label{eq:steering}
\end{equation}
Stacking the responses in \eqref{eq:steering} gives the unit-norm array vector $\mathbf a(\mathbf z(t))$. The corresponding downlink channel is
\begin{equation}
\mathbf h(t)=\sqrt{\beta_M}\,\mathbf a(\mathbf z(t)),
\label{eq:downlink_channel}
\end{equation}
where $\beta_M>0$ is the channel power gain. The phase of each element is measured relative to the propagation phase at element $1$. Subtracting $r_1$ sets the first element phase to zero and preserves the phase differences across the array. The removed common phase does not affect received power or SNR at an individual instant. We use this same phase convention for motion inference and coherent beam synthesis throughout the service interval. In the primary system benchmarks, $\beta_M$ is held constant to isolate spatial alignment from range variation, whereas Section~\ref{sec:results} additionally evaluates range-dependent path loss and near-field Rician multipath.

The BS transmits a normalized data symbol $s(t)$ using a power-constrained beamforming vector $\mathbf f\in\mathbb C^M$ and a transmit-power budget $P$. The received signal is
\begin{equation}
y(t)=\sqrt P\,\mathbf h^{\mathsf H}(t)\mathbf f\,s(t)+n(t),
\label{eq:received_signal}
\end{equation}
where $n(t)\sim\mathcal{CN}(0,\sigma^2)$ is independent receiver noise. With reference SNR $\gamma_M=P\beta_M/\sigma^2$, the instantaneous Shannon spectral efficiency (SE) in bit/s/Hz is
\begin{equation}
R(t,\mathbf f)=\log_2\!\left(1+\gamma_M\left|\mathbf a^{\mathsf H}(\mathbf z(t))\mathbf f\right|^2\right).
\label{eq:instantaneous_se}
\end{equation}
The instantaneous matched beam $\mathbf f=\mathbf a(\mathbf z(t))$ achieves unit overlap and reference rate $\log_2(1+\gamma_M)$, whereas near-field point focusing (\textbf{\mbox{NF-Point}}) holds a beam aimed at one predicted position and experiences time-varying rate under motion.

\subsection{Finite Updates and Cross-Interval Beam Service}
\label{sec:motion_focusing}
Because observation arrivals, motion inference, beam computation, and array reconfiguration incur finite latency, we model the beam controller with discrete decisions and delayed execution. At each decision time $t_k$, the controller selects target vector $\mathbf f_{b,k}$ solely from arrived observations. As shown in Fig.~\ref{fig:fig2}, the update begins at $t_a=t_k+T_{\rm d}$ and governs the service interval
\begin{equation}
\mathcal H_k=[t_k+T_{\rm d},\ t_k+T_{\rm d}+T_{\rm h}],
\label{eq:hold}
\end{equation}
with $t_b=t_a+T_{\rm h}$. Moreover, practical analog or hybrid weight networks, movable elements, port switches, and liquid-crystal phase shifters settle over a finite reconfiguration time constant $\tau_{\rm r}\ge 0$~\cite{Wang2026MAMovement,Wang2024LiquidCrystal,Dinis2026FASSwitching,Ting2026LCRIS}. Across consecutive intervals, the realized unit-power transmit vector on $\mathcal H_k$ relaxes from $\mathbf f_{b,k-1}$ toward $\mathbf f_{b,k}$ according to
\begin{equation}
\tilde{\mathbf f}_k(t)=\frac{(1-\alpha_k(t))\mathbf f_{b,k-1}+\alpha_k(t)\mathbf f_{b,k}}{\bigl\|(1-\alpha_k(t))\mathbf f_{b,k-1}+\alpha_k(t)\mathbf f_{b,k}\bigr\|_2},
\label{eq:point_gain}
\end{equation}
where $\alpha_k(t)=1-\exp(-(t-t_a)/\tau_{\rm r})$ for $\tau_{\rm r}>0$, recovering instantaneous step switching $\alpha_k(t)\equiv 1$ as $\tau_{\rm r}\to 0^+$. The beam $\mathbf f_{b,k}=\mathbf a(\widehat{\mathbf z}_\star)$ focuses on the predicted position at $t_\star=t_a+T_{\rm h}/2$ and can lose gain as the user moves away. Unaligned consecutive beams can also interfere destructively near $t_a$. Substituting $\tilde{\mathbf f}_k(t)$ into the time-averaged interval SE gives
\begin{equation}
\overline R_k=\frac{1}{T_{\rm h}}\int_{\mathcal H_k}
R\!\left(t,\tilde{\mathbf f}_k(t)\right)dt.
\label{eq:se}
\end{equation}
Interval performance therefore depends on the positions visited, the residence time near each location, and phase continuity across interval boundaries.

\subsection{Available Observations and Causal Beam Shaping}
\label{sec:shaping_task}

To forecast motion over $\mathcal H_k$, the controller combines high-rate uplink communication pilots with sparser sensing reports that may arrive with delay. Pilot-based geometric accuracy depends on the pilot SNR, while sensing reports constrain motion at individual times or over intervals.

The single antenna user transmits known uplink pilots to the BS. At global pilot time $t_i$, the normalized observation across all $M$ elements is
\begin{equation}
\mathbf y_i=\alpha_i\mathbf a(\mathbf z(t_i))+\mathbf n_i,
\label{eq:pilot}
\end{equation}
where $\alpha_i$ is an unknown complex gain accounting for pilot strength and common propagation phase, and $\mathbf n_i\sim\mathcal{CN}(\mathbf0,\sigma_{\rm p}^2\mathbf I_M)$ is receiver noise. The array response links the pilot to the user's geometry at its acquisition time. The BS retains the full array observation before spatial combining, and pilot acquisitions follow a global clock independently of the held downlink beam.

\begin{figure}[!t]
\centering
\includegraphics[width=0.96\columnwidth]{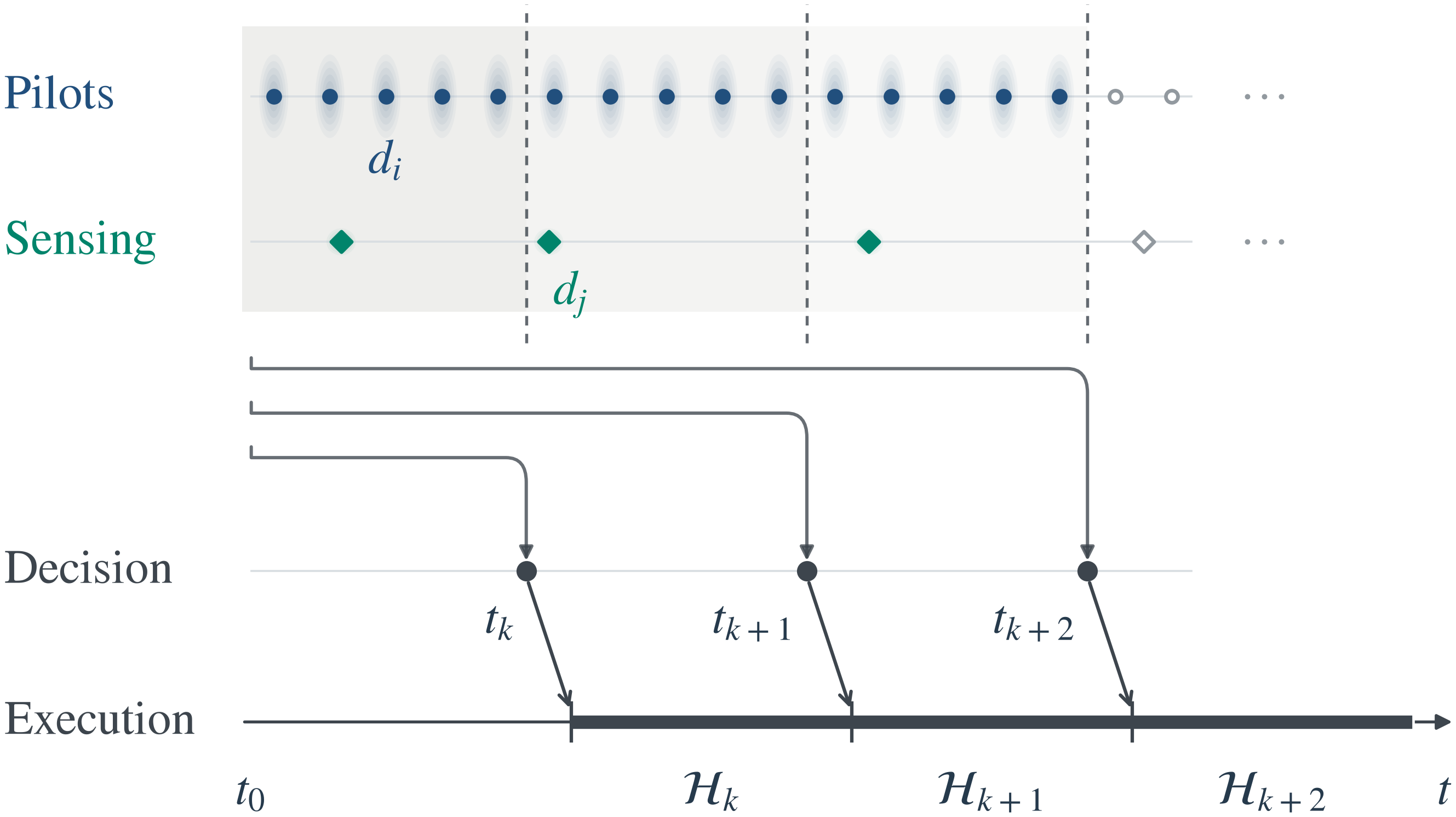}
\caption{Observation arrivals across successive beam decisions.}
\label{fig:fig3}
\end{figure}

Sensing reports provide geometric information about the same motion on point or interval supports. For report $j$ with support $[a_j,b_j]$, the 2D Cartesian measurement model is
\begin{equation}
\bar{\mathbf z}_j^{\rm s}=\mathcal L_j[\mathbf z]+\boldsymbol\epsilon_j,
\label{eq:temporal}
\end{equation}
where $\boldsymbol\epsilon_j\sim\mathcal N(\mathbf0,\mathbf R_j^{\rm s})$ is the report error and the known linear operator $\mathcal L_j$ specifies the sensed quantity and its support. Representative reports include position fixes, independently measured displacement, and time averaged position. For example, a global navigation satellite system (GNSS) fix can be converted to the array's local Cartesian frame, odometry can provide displacement, and a localization frontend can report its time averaged position estimates. Their operators are
\begin{equation}
\mathcal L_j[\mathbf z]=
\begin{cases}
\mathbf z(b_j), & \text{position},\\
\mathbf z(b_j)-\mathbf z(a_j), & \text{displacement},\\
\displaystyle\frac{1}{b_j-a_j}\int_{a_j}^{b_j}\mathbf z(t)\,dt, & \text{time average}.
\end{cases}
\label{eq:operators}
\end{equation}
Position reports have point support $a_j=b_j$, whereas interval reports use $b_j>a_j$. The model also accommodates other known linear temporal operators. Measurement noises are independent across sources. Figure~\ref{fig:fig3} summarizes the complementary reporting characteristics of pilots and sensing and shows which arrivals can inform successive beam decisions.

Each observation record $o_i$ carries a unique identity, measured value, noise covariance, acquisition support $[a_i,b_i]$ ($a_i=b_i$ for pilots), and arrival time $d_i\ge b_i$, where $d_i-b_i$ is its reporting delay (distinct from the execution delay $T_{\rm d}$). At decision time $t_k$, the causal information set is
\begin{equation}
\mathcal F_k=\{o_i\mid d_i\le t_k\},
\label{eq:info_set}
\end{equation}
indexed by $\mathcal P_k$ for arrived pilots and $\mathcal S_k$ for arrived sensing reports. As shown in Fig.~\ref{fig:fig3}, records with $d_i>t_k$ are excluded from decision $k$ even if $d_i<t_a$, and enter only subsequent decisions on their native supports $[a_i,b_i]$.

Using the available information $\mathcal F_k$, the controller must select one beam to serve the entire future interval $\mathcal H_k$. Let $\overline R_k(\mathbf f)$ denote the interval average SE in \eqref{eq:se} when a candidate beam $\mathbf f$ is held throughout that interval. Since future user motion is unknown at $t_k$, an ideal causal design maximizes the conditional expected interval SE:
\begin{equation}
\mathbf f_{b,k}^{\star}\in
\underset{\|\mathbf f\|_2^2\le1}{\arg\max}\,
\mathbb E\!\left[\overline R_k(\mathbf f)\mid\mathcal F_k\right].
\label{eq:beam_power}
\end{equation}
The expectation is over future user motion given $\mathcal F_k$ and an assumed predictive model, yielding the single-column beam $\mathbf F_{b,k}=\mathbf f_{b,k}$. Because the observation models alone do not fix this predictive distribution, we adopt a two-stage causal design. First, a motion predictor maps $\mathcal F_k$ to a forecast over the service interval:
\begin{equation}
\widetilde{\mathbf z}_k(t)=\Phi_k(t,\mathcal F_k),\quad t\in\mathcal H_k,
\label{eq:motion_inference_task}
\end{equation}
by inferring a degree-$p$ continuous historical entity with $p\ge1$ and smoothness order $1\le m\le p$, then extrapolating its terminal motion to order $0\le N\le p$. Sections~\ref{sec:inference}--\ref{sec:beam} detail these steps. Second, over the cross-window support $\widetilde{\mathcal H}_k=[t_a,t_b+\Delta_{\rm t}]$ with $\Delta_{\rm t}=\min(2\tau_{\rm r},T_{\rm h})$ and normalized activation kernel $w_k(t)\ge0$, $\int_{\widetilde{\mathcal H}_k}w_k(t)\,dt=1$, the controller selects a beam maximizing weighted coherent alignment
\begin{equation}
\mathbf f_{b,k}\in\underset{\|\mathbf f\|_2^2\le1}{\arg\max}\,
\left|\int_{\widetilde{\mathcal H}_k}
w_k(t)\,\mathbf a^{\mathsf H}(\widetilde{\mathbf z}_k(t))\mathbf f\,dt\right|^2,
\label{eq:interval_beam_task}
\end{equation}
fixing its global phase to align constructively with $\mathbf f_{b,k-1}$ at $\widetilde{\mathbf z}_k(t_a)$. Together, the two stages use the observations available at $t_k$ to infer the motion history, predict the trajectory, and synthesize one phase-continuous beam per interval. The coherent objective is a tractable mean-forecast surrogate and does not generally solve the conditional expected-SE problem in \eqref{eq:beam_power}. The realized future trajectory is used only to evaluate the resulting beam.

\section{Geometric Entity Inference and Pilot Refinement}
\label{sec:inference}
To realize the causal predictor $\Phi_k$ in \eqref{eq:motion_inference_task} at $t_k$, the controller fuses arrived pilots $\mathcal P_k$ and potentially delayed sensing reports $\mathcal S_k$, whose supports $[a_i,b_i]$ lie within $[t_0,t_k]$ since $b_i\le d_i\le t_k$. This section estimates a continuous historical entity on $[t_0,t_k]$ that couples all arrived records on their native supports and supplies the terminal readout at $t_k$.

\subsection{Historical Geometric Entity for Motion Inference}
\label{sec:geometric_entity}
Following~\cite{Fan2026GeometricTracking}, we model user motion as a continuous physical-time trajectory over $[t_0,t_k]$, jointly capturing path geometry and traversal speed that govern interval sensing \eqref{eq:operators} and held-beam residence weighting. Each arrived report in $\mathcal S_k$ constrains the entity on its acquisition support $[a_j,b_j]\subset[t_0,t_k]$. Pilots in $\mathcal P_k$ constrain positions at their own timestamps. Both enter the coefficient fit from which the terminal readout is taken at $t_k$.

We parameterize the trajectory over $[t_0,t_k]$ with a degree-$p$ B-spline~\cite{deBoor2001} on a nondecreasing knot sequence $\tau_0,\ldots,\tau_{K_k+p}$ clamped at $t_0$ and $t_k$ with simple interior knots. Stacking the $K_k\ge p+1$ planar control points into $\mathbf c=[\mathbf c_0^{\mathsf T},\ldots,\mathbf c_{K_k-1}^{\mathsf T}]^{\mathsf T}\in\mathbb R^{2K_k}$ expresses the candidate entity as
\begin{equation}
\mathcal G_k(t,\mathbf c)=\sum_{\ell=0}^{K_k-1}b_{\ell,p}(t)\mathbf c_\ell=\mathbf B_k(t)\mathbf c,
\label{eq:geometric_entity}
\end{equation}
where $\mathbf B_k(t)=\bigl[b_{0,p}(t)\mathbf I_2,\ldots,b_{K_k-1,p}(t)\mathbf I_2\bigr]\in\mathbb R^{2\times2K_k}$. Fixing the knots at $t_k$ reduces continuous trajectory inference to estimating $\mathbf c$. Starting from
\begin{equation}
b_{\ell,0}(t)=
\begin{cases}
1, & \tau_\ell\le t<\tau_{\ell+1},\\
0, & \text{otherwise},
\end{cases}
\label{eq:spline_initial_basis}
\end{equation}
the Cox--de Boor recursion generates degrees $q=1,\ldots,p$ via
\begin{equation}
b_{\ell,q}(t)=\frac{t-\tau_\ell}{\tau_{\ell+q}-\tau_\ell}b_{\ell,q-1}(t)
+\frac{\tau_{\ell+q+1}-t}{\tau_{\ell+q+1}-\tau_{\ell+1}}b_{\ell+1,q-1}(t),
\label{eq:spline_recursion}
\end{equation}
with zero-denominator terms vanishing and one-sided limits at $t_0$ and $t_k$. Each basis function spans at most $p+1$ adjacent knot intervals with $C^{p-1}$ interior continuity. Because $\mathcal L[\mathcal G_k]=\mathcal L[\mathbf B_k]\mathbf c$ for every linear operator in \eqref{eq:operators}, point values, endpoint differences, interval averages, and terminal derivatives $\mathbf B_k^{(n)}(t_k)\mathbf c$ are evaluated directly from the basis, even when true motion does not lie in the spline space.

\subsection{Coupling Communication and Sensing Observations}
\label{sec:observation_coupling}
Each candidate entity in \eqref{eq:geometric_entity} determines the geometric quantities measured by sensing reports on their acquisition supports and the array responses associated with pilots at their acquisition times. We evaluate its agreement with both types of observation through the shared control vector $\mathbf c$, giving the joint observation cost
\begin{equation}
\mathcal J_k(\mathbf c)=\sum_{j\in\mathcal S_k}\|\mathbf A_{j,k}\mathbf c-\bar{\mathbf z}_j^{\rm s}\|_{(\mathbf R_j^{\rm s})^{-1}}^2+2\sum_{i\in\mathcal I_k}w_i\ell_i\!\left(\mathbf B_k(t_i)\mathbf c\right).
\label{eq:common_factor}
\end{equation}
The first sum compares the entity prediction $\mathbf A_{j,k}\mathbf c$ with the sensing report $\bar{\mathbf z}_j^{\rm s}$, weighted by the inverse of its error covariance $\mathbf R_j^{\rm s}$. The second uses the gain-profiled negative log likelihood $\ell_i$ of the raw pilot $\mathbf y_i$ to assess the array response at the entity position $\mathbf B_k(t_i)\mathbf c$. Here $\mathcal I_k\subseteq\mathcal P_k$ denotes the accepted pilots, and $w_i>0$ is the weight fixed for the current decision.

For sensing report $j$, applying its temporal operator to the entity gives
\begin{equation}
\bar{\mathbf z}_j^{\rm s}=\mathbf A_{j,k}\mathbf c+\boldsymbol\epsilon_j,
\label{eq:operator_matrix}
\end{equation}
where $\mathbf A_{j,k}=\mathcal L_j[\mathbf B_k]$ maps the controls to the measured geometric quantity on $[a_j,b_j]$ via \eqref{eq:operators}. Under the chosen entity model and Gaussian report error, the first sum in \eqref{eq:common_factor} is the doubled sensing negative log likelihood up to a constant independent of $\mathbf c$.

A pilot contributes the complex array observation $\mathbf y_i$ in \eqref{eq:pilot}. For a candidate position $\mathbf z$, eliminating the unknown gain using $\widehat\alpha_i(\mathbf z)=\mathbf a^{\mathsf H}(\mathbf z)\mathbf y_i$ \cite{GolubPereyra1973} gives the profiled cost
\begin{equation}
\ell_i(\mathbf z)=\frac{\|\mathbf y_i\|_2^2-|\mathbf a^{\mathsf H}(\mathbf z)\mathbf y_i|^2}{\sigma_{\rm p}^2}.
\label{eq:profile}
\end{equation}
Position is the intermediate model variable $\mathbf z=\mathbf B_k(t_i)\mathbf c$ through which the entity predicts the array response. Thus, the pilot term in \eqref{eq:common_factor} retains the raw observation's likelihood as a function of the controls. Its factor of two matches the scale of the real Gaussian sensing residuals.

The raw pilot cost in \eqref{eq:profile} can have multiple spatial minima, since different positions may produce similar array responses. To select a local expansion point for pilot $i$, we first use the sensing reports available at $t_k$ to build a historical motion guide and evaluate it at the pilot time $t_i$. A bounded search combines a robust pilot cost with a penalty for departure from the guide to find an association position $\bar{\mathbf z}_i$. The guide affects association and weighting, but its penalty is not added as an independent pilot factor. Residual reduction, consistency with the guide, and information conditioning determine the accepted pilot set $\mathcal I_k$ and weights $w_i$, which remain fixed during the current decision.

At $\bar{\mathbf z}_i$, let $\mathbf g_i$ and $\mathbf I_i$ be the score and projected Gauss--Newton information in \eqref{eq:raw_jacobian}. With $\boldsymbol\delta_i=\mathbf B_k(t_i)\mathbf c-\bar{\mathbf z}_i$, the local pilot model is
\begin{equation}
2w_i\ell_i\!\left(\mathbf B_k(t_i)\mathbf c\right)\simeq 2w_i\mathbf g_i^{\mathsf T}\boldsymbol\delta_i+w_i\boldsymbol\delta_i^{\mathsf T}\mathbf I_i\boldsymbol\delta_i+\xi_i,
\label{eq:pilot_quadratic}
\end{equation}
where $\xi_i=2w_i\ell_i(\bar{\mathbf z}_i)$ is independent of $\mathbf c$. The score and information enter the entity fit, with $\mathbf I_i$ contributing $w_i\mathbf B_k^{\mathsf T}(t_i)\mathbf I_i\mathbf B_k(t_i)$ to the coefficient precision. Proposition~\ref{prop:pilot_information}, proved in Appendix~\ref{app:pilot_information}, identifies the directions resolved by a pilot with unknown gain.

\begin{proposition}[Pilot-Resolved Directions]
\label{prop:pilot_information}
For $\widehat\alpha_i(\bar{\mathbf z}_i)\ne0$, let $\mathbf a_i=\mathbf a(\bar{\mathbf z}_i)$. A real position perturbation $\mathbf v\in\mathbb R^2$ has zero unregularized projected pilot information exactly when its first-order steering change can be absorbed into the unknown gain:
\begin{equation}
\mathbf v^{\mathsf T}\mathbf I_i\mathbf v=0
\quad\Longleftrightarrow\quad
(\mathbf I_M-\mathbf a_i\mathbf a_i^{\mathsf H})
\left.\frac{\partial\mathbf a}{\partial\mathbf z^{\mathsf T}}\right|_{\bar{\mathbf z}_i}\mathbf v=\mathbf0.
\label{eq:pilot_information_null}
\end{equation}
\end{proposition}

The acceptance test retains well-conditioned pilot factors, while sensing constrains the same entity on its own supports.

\begin{figure}[!t]
\centering
\includegraphics[width=0.84\columnwidth]{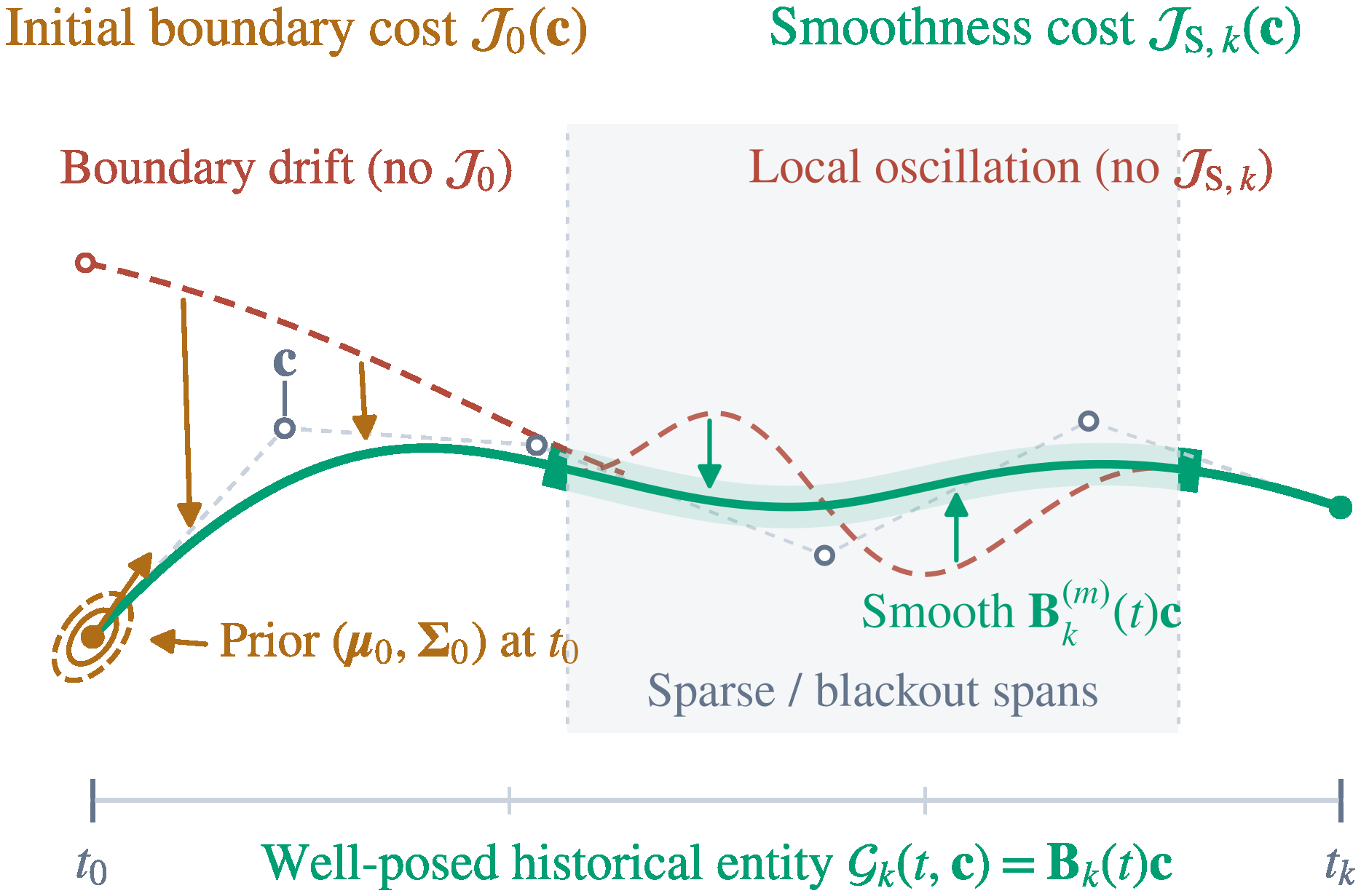}
\caption{Boundary anchoring and smoothing of the historical entity.}
\label{fig:fig4}
\end{figure}

\subsection{Joint Entity Inference and Pilot Refinement}
\label{sec:joint_inference}
Because each B-spline basis function in \eqref{eq:geometric_entity} has compact support over at most $p+1$ knot spans, minimizing the observation cost alone cannot guarantee a well-posed trajectory across $[t_0,t_k]$. When observations arrive sparsely, experience blackouts, or provide only differential or directional constraints, the control points can be weakly constrained at the initial boundary $t_0$ and across interior knot spans, as illustrated in Fig.~\ref{fig:fig4}. We therefore combine the quadratic observation cost $\widehat{\mathcal J}_k(\mathbf c)$ with an initial boundary cost $\mathcal J_0(\mathbf c)$ and a smoothness cost $\mathcal J_{{\rm S},k}(\mathbf c)$ by solving
\begin{equation}
\widehat{\mathbf c}_{k|k}=\arg\min_{\mathbf c}\;\widehat{\mathcal J}_k(\mathbf c)+\mathcal J_0(\mathbf c)+\mathcal J_{{\rm S},k}(\mathbf c),
\label{eq:entity_objective}
\end{equation}
where $\widehat{\mathcal J}_k(\mathbf c)$ is obtained by substituting \eqref{eq:pilot_quadratic} into \eqref{eq:common_factor} without $\xi_i$.

The two regularizing costs in \eqref{eq:entity_objective} stabilize the initial boundary and weakly observed parts of the history. When early observations provide insufficient information about the initial motion, the boundary cost $\mathcal J_0(\mathbf c)$ anchors the derivatives of orders $0,\ldots,p$ at $t_0$ to a fixed Gaussian prior $(\boldsymbol\mu_0,\boldsymbol\Sigma_0)$ with $\boldsymbol\Sigma_0\succ0$ through
\begin{equation}
\mathcal J_0(\mathbf c)=\|\mathbf D_0\mathbf c-\boldsymbol\mu_0\|_{\boldsymbol\Sigma_0^{-1}}^2,
\label{eq:init_cost}
\end{equation}
where $\mathbf D_0$ stacks the corresponding basis derivatives from the right at $t_0$. During position-report blackouts or sparse intervals, limited observation information can also permit spurious variation across interior knot spans. The smoothness cost $\mathcal J_{{\rm S},k}(\mathbf c)$ discourages such variation by penalizing the $m$th derivative over $[t_0,t_k]$ with weight $\lambda_{\rm S}>0$ as
\begin{equation}
\mathcal J_{{\rm S},k}(\mathbf c)=\lambda_{\rm S}\int_{t_0}^{t_k}\|\mathbf B_k^{(m)}(t)\mathbf c\|_2^2\,dt.
\label{eq:reg_cost}
\end{equation}
The degree $p$ sets the representation capacity, while $m$ specifies the derivative penalized for smoothness. Both the smoothness integral and sensing averages are evaluated over the polynomial knot spans, with numerical settings given in Section~\ref{sec:protocol}.

In \eqref{eq:pilot_quadratic}, each pilot model is initially expanded around the sensing-guided point $\bar{\mathbf z}_i$. Solving \eqref{eq:entity_objective} with these initial models yields an initial entity $\widehat{\mathbf c}^{(0)}_{k|k}$ whose trajectory pools all arrived sensing reports and pilot likelihoods across $[t_0,t_k]$. To refine the local pilot approximations using this joint trajectory estimate, we query the initial entity at each accepted pilot timestamp to obtain
\begin{equation}
\widehat{\mathbf z}^{(0)}_k(t_i)=\mathbf B_k(t_i)\widehat{\mathbf c}^{(0)}_{k|k},\quad i\in\mathcal I_k.
\label{eq:feedback_seed}
\end{equation}
Each query $\widehat{\mathbf z}^{(0)}_k(t_i)$ seeds one bounded local refinement of the raw pilot cost \eqref{eq:profile} to update the linearization point, gradient $\mathbf g_i$, and information $\mathbf I_i$ in \eqref{eq:pilot_quadratic}, while keeping $\mathcal I_k$ and $w_i$ fixed. Valid updates replace the corresponding pilot models in $\widehat{\mathcal J}_k(\mathbf c)$, whereas ill-conditioned updates retain their initial expansions. Solving \eqref{eq:entity_objective} a second time from the same initial cost $\mathcal J_0(\mathbf c)$ then yields the final estimate $\widehat{\mathbf c}_{k|k}$. Because refined pilot models replace their initial expansions and each decision refits from $\mathcal J_0(\mathbf c)$, every arrived observation is counted once without accumulating redundant prior precision, as summarized in Algorithm~\ref{alg:main}.

Since all three terms in \eqref{eq:entity_objective} are quadratic in $\mathbf c$, each fit reduces to a whitened linear least-squares problem. With $\mathbf H_k$ denoting the stacked whitened design matrix of the final fit, column-scaled QR factorization yields $\widehat{\mathbf c}_{k|k}$ together with the coefficient covariance
\begin{equation}
\mathbf P_{k|k}=(\mathbf H_k^{\mathsf T}\mathbf H_k)^{-1}.
\label{eq:entity_cov}
\end{equation}
The pair $(\widehat{\mathbf c}_{k|k},\mathbf P_{k|k})$ characterizes the complete historical entity over $[t_0,t_k]$, defining the mean trajectory $\widehat{\mathbf z}_k(t)=\mathbf B_k(t)\widehat{\mathbf c}_{k|k}$ and the inverse-information covariance induced by the boundary prior, smoothness regularization, and local observation precision.

For the prediction interface at the decision boundary $t_k$, let $\widehat{\mathbf z}_k^{(n)}(t_k)=\mathbf B_k^{(n)}(t_k)\widehat{\mathbf c}_{k|k}\in\mathbb R^2$ denote the $n$th-order terminal motion derivative. Orders $n=0$, $n=1$, and $n=2$ correspond to position, velocity, and acceleration, respectively. Stacking these derivatives up to order $N$ yields the terminal derivative state, also called the $N$-jet vector,
\begin{equation}
\widehat{\boldsymbol j}_{N,k}
=\bigl[\widehat{\mathbf z}_k^{(0)}(t_k)^{\mathsf T},\,\ldots,\,\widehat{\mathbf z}_k^{(N)}(t_k)^{\mathsf T}\bigr]^{\mathsf T}
=\mathbf D_{N,k}\widehat{\mathbf c}_{k|k},
\label{eq:jet}
\end{equation}
where $\mathbf D_{N,k}=\bigl[\mathbf B_k(t_k)^{\mathsf T},\,\ldots,\,\mathbf B_k^{(N)}(t_k)^{\mathsf T}\bigr]^{\mathsf T}\in\mathbb R^{2(N+1)\times 2K_k}$ is the terminal readout matrix, with covariance
\begin{equation}
\boldsymbol\Sigma_{N,k}=\mathbf D_{N,k}\mathbf P_{k|k}\mathbf D_{N,k}^{\mathsf T}.
\label{eq:jet_cov}
\end{equation}

Proposition~\ref{prop:fusion_forecast}, whose proof is in Appendix~\ref{app:fusion_proof}, links forecast uncertainty through \eqref{eq:jet_cov} to accepted pilots ($\mathrm P$) and arrived interval reports ($\mathrm S$). The fits share point reports and positive-definite quadratic regularization on a finite-dimensional linear trajectory space, with fixed pilot factors and independent errors.

\begin{proposition}[Complementary Forecast Information from Asynchronous Pilots and Sensing]
\label{prop:fusion_forecast}
Fusion lowers modeled forecast variance below both single-source values exactly when each source measures a component still correlated with the forecast given the other. For any supported Taylor order $N$ ($0\le N\le p$ for the spline), let $y$ be a directional position forecast at any time in the activation window. Write $\sigma_r^2=\operatorname{Var}_r(y)$ and $\boldsymbol\kappa_{r,N}(s)=\operatorname{Cov}_r(\mathbf z(s),y)$ in fit $r$. Then
\begin{align}
&\sigma_{\rm P+S}^2<\min\{\sigma_{\rm P}^2,\sigma_{\rm S}^2\}
\quad\Longleftrightarrow\quad
\mathcal C_{\rm P\mid S}\land\mathcal C_{\rm S\mid P}.
\label{eq:fusion_forecast_precision}\\
\intertext{Condition $\mathcal C_{\rm P\mid S}$ asks whether an array-resolved pilot component lowers the interval fit's variance $\sigma_{\rm S}^2$. Condition $\mathcal C_{\rm S\mid P}$ asks whether a displacement or interval mean lowers the pilot fit's variance $\sigma_{\rm P}^2$.}
&\quad\mathcal C_{\rm P\mid S}\colon\quad
\exists\,i\in\mathcal I_k:\ \mathbf I_i^{1/2}\boldsymbol\kappa_{\rm S,N}(t_i)\ne\mathbf0,\notag\\
&\quad\mathcal C_{\rm S\mid P}\colon\quad
\exists\,j\in\mathcal S_k:\ a_j<b_j\ \land\
\mathcal L_j[\boldsymbol\kappa_{\rm P,N}]\ne\mathbf0.\notag
\end{align}
\end{proposition}

For noiseless one-dimensional affine motion, a pilot can anchor the forecast at one endpoint, while the interval displacement sets its slope. More generally, the geometric entity connects pilot samples and interval reports on their different time supports to a common forecast readout. When both conditions hold, these sources contribute complementary information about the selected forecast component. The fused entity's mean forecast then guides beam shaping over the service interval.

\begin{algorithm}[!t]
\caption{Causal Entity Inference and Beam Shaping}
\label{alg:main}
\begin{algorithmic}[1]
\Require $\mathcal P_k$, $\mathcal S_k$, and previous beam $\mathbf f_{b,k-1}$.
\Ensure Entity $(\widehat{\mathbf c}_{k|k},\mathbf P_{k|k})$ and beamformer $\mathbf F_{b,k}$.
\Statex \textbf{Stage I: Initial Historical Entity Fit}
\State Form sensing factors $\{(\mathbf A_{j,k},\bar{\mathbf z}_j^{\rm s},\mathbf R_j^{\rm s})\}_{j\in\mathcal S_k}$ via \eqref{eq:operator_matrix}.
\State Screen accepted pilot set $\mathcal I_k$ via \eqref{eq:profile}.
\State Form initial pilot factors $\{(w_i,\bar{\mathbf z}_i,\mathbf g_i,\mathbf I_i)\}_{i\in\mathcal I_k}$ via \eqref{eq:pilot_quadratic}.
\State Solve \eqref{eq:entity_objective} from $\mathcal J_0(\mathbf c)$ for initial controls $\widehat{\mathbf c}_{k|k}^{(0)}$.
\Statex \textbf{Stage II: Single-Pass Pilot Refinement and Refit}
\For{each accepted pilot $i\in\mathcal I_k$}
    \State Query entity position $\widehat{\mathbf z}_k^{(0)}(t_i)$ via \eqref{eq:feedback_seed}.
    \State Refine pilot expansion around $\widehat{\mathbf z}_k^{(0)}(t_i)$ via \eqref{eq:profile}.
    \State Update $(\bar{\mathbf z}_i,\mathbf g_i,\mathbf I_i)$ in \eqref{eq:pilot_quadratic} if well-conditioned.
\EndFor
\State Re-solve \eqref{eq:entity_objective} from $\mathcal J_0(\mathbf c)$ for refined controls $\widehat{\mathbf c}_{k|k}$.
\State Compute entity covariance $\mathbf P_{k|k}$ via \eqref{eq:entity_cov}.
\Statex \textbf{Stage III: Terminal Readout and Cross-Window GPBS}
\State Extract terminal $N$-jet $\widehat{\boldsymbol j}_{N,k}$ via \eqref{eq:jet}.
\State Evaluate trajectory forecast $\widetilde{\mathbf z}_k(t)$ over $\widetilde{\mathcal H}_k$ via \eqref{eq:taylor_forecast}.
\State Integrate cross-window vector $\mathbf q_k$ over $\widetilde{\mathcal H}_k$ via \eqref{eq:q}.
\State Compute handoff phase $\vartheta_k$ at $\widetilde{\mathbf z}_k(t_a)$.
\State Synthesize held beamformer $\mathbf F_{b,k}$ via \eqref{eq:coherent}.
\end{algorithmic}
\end{algorithm}

\section{Prediction and Coherent Beam Shaping}
\label{sec:beam}
The terminal jet from the refitted entity specifies the predicted positions on the activation window $\widetilde{\mathcal H}_k$. Integrating their spherical array responses in physical time gives the held beam in \eqref{eq:interval_beam_task}.

\subsection{Terminal Motion Prediction and Its Geometry}
\label{sec:geometry_readout}
Because the B-spline basis $\mathbf B_k(t)$ is supported only on $[t_0,t_k]$, predicting the future user trajectory $\widetilde{\mathbf z}_k(t)\in\mathbb R^2$ over $\widetilde{\mathcal H}_k$ uses the terminal derivatives $\widehat{\mathbf z}_k^{(n)}(t_k)$ stacked in $\widehat{\boldsymbol j}_{N,k}$. With the lookahead offset $\tau=t-t_k$, the $N$th-order Taylor trajectory forecast is
\begin{equation}
\widetilde{\mathbf z}_k(t_k+\tau)=
\sum_{n=0}^{N}\frac{\tau^n}{n!}\widehat{\mathbf z}_k^{(n)}(t_k)
=\mathbf E_N(\tau)\widehat{\boldsymbol j}_{N,k},
\label{eq:taylor_forecast}
\end{equation}
where the temporal propagation matrix is given by
\begin{equation}
\mathbf E_N(\tau)=\left[\mathbf I_2,\,\tau\mathbf I_2,\,\ldots,\,\frac{\tau^N}{N!}\mathbf I_2\right]\in\mathbb R^{2\times 2(N+1)}.
\label{eq:e_prop}
\end{equation}
Equations~\eqref{eq:taylor_forecast}--\eqref{eq:e_prop} separate the forecast order from the historical fit. For a fixed degree-$p$ entity, $\widehat{\boldsymbol j}_{N,k}$ can be formed at any supported order without refitting the historical trajectory. The resulting forecasts range from constant position ($N=0$) and constant velocity ($N=1$) to constant acceleration ($N=2$), with higher orders retaining additional terminal derivatives.

In practice, selecting $N$ balances expressiveness against noise amplification across $\mathcal H_k$. Within the inference model, the forecast covariance at offset $\tau$ is
\begin{equation}
\operatorname{Cov}\!\left(\widetilde{\mathbf z}_k(t_k+\tau)\right)
=\mathbf E_N(\tau)\boldsymbol\Sigma_{N,k}\mathbf E_N^{\mathsf T}(\tau),
\label{eq:forecast_cov}
\end{equation}
where $\tau^n/n!$ amplifies variance in high-order boundary derivatives under sparse or noisy observations. Choosing $N=2$ truncates noise-sensitive higher orders while capturing curved near-field maneuvers.

\subsection{Cross-Window Coherent Beam Shaping in Physical Time}
\label{sec:coherent_beam}
For $\tau_{\rm r}>0$, the kernel $w_k$ in \eqref{eq:interval_beam_task} is obtained by normalizing the weights $\alpha_k(t)$ on $\mathcal H_k$ and $\alpha_k(t_b)e^{-(t-t_b)/\tau_{\rm r}}$ on $\widetilde{\mathcal H}_k\setminus\mathcal H_k$. As $\tau_{\rm r}\to0^+$, it approaches uniform boxcar weighting. GPBS integrates the predicted responses as
\begin{equation}
\mathbf q_k=\int_{\widetilde{\mathcal H}_k}
w_k(t)\,\mathbf a(\widetilde{\mathbf z}_k(t))\,dt.
\label{eq:q}
\end{equation}
All responses use the first-element phase reference in \eqref{eq:steering}. Time integration weights each visit by its duration. Even a stationary dwell contributes its response over the full dwell time.

Define the weighted spatial correlation of the predicted responses as
\begin{equation}
\mathbf C_k=\int_{\widetilde{\mathcal H}_k}
w_k(t)\,\mathbf a(\widetilde{\mathbf z}_k(t))\mathbf a^{\mathsf H}(\widetilde{\mathbf z}_k(t))\,dt.
\label{eq:cov_beam}
\end{equation}
The first-element reference gives $\mathbf q_k=\sqrt M\,\mathbf C_k\mathbf e_1$, where $\mathbf e_1$ is the first canonical basis vector, and $\mathbf C_k-\mathbf q_k\mathbf q_k^{\mathsf H}\succeq\mathbf 0$ since $w_k$ is nonnegative and normalized. For any feasible beam, $0\le|\mathbf a^{\mathsf H}\mathbf f|^2\le1$ and concavity $\log_2(1+\gamma g)\ge g\log_2(1+\gamma)$ for $\gamma\ge0$ yield
\begin{equation}
\begin{aligned}
&\int_{\widetilde{\mathcal H}_k} w_k(t)
\log_2\!\left(1+\gamma\left|
\mathbf a^{\mathsf H}(\widetilde{\mathbf z}_k(t))\mathbf f
\right|^2\right)\,dt\\
&\qquad\ge\log_2(1+\gamma)\,\mathbf f^{\mathsf H}\mathbf C_k\mathbf f\\
&\qquad\ge\log_2(1+\gamma)\,|\mathbf q_k^{\mathsf H}\mathbf f|^2.
\end{aligned}
\label{eq:coherent_rate_bound}
\end{equation}
Thus \eqref{eq:interval_beam_task} maximizes an explicit lower bound on the kernel-weighted SE along $\widetilde{\mathbf z}_k(t)$, without establishing conditional expected-SE optimality in \eqref{eq:beam_power}.

By the Cauchy--Schwarz inequality, $|\mathbf q_k^{\mathsf H}\mathbf f|^2\le\|\mathbf q_k\|_2^2\|\mathbf f\|_2^2$ is maximized by every unit-norm rotation of $\mathbf q_k$. Even with a common first-element phase reference, consecutive beams can arrive out of phase at the predicted handoff position $\widetilde{\mathbf z}_k(t_a)$ and interfere destructively during \eqref{eq:point_gain}. For a nonzero product, the handoff phase is
\begin{equation}
\vartheta_k=\arg\!\left[
\bigl(\mathbf a^{\mathsf H}(\widetilde{\mathbf z}_k(t_a))\mathbf f_{b,k-1}\bigr)
\bigl(\mathbf q_k^{\mathsf H}\mathbf a(\widetilde{\mathbf z}_k(t_a))\bigr)
\right].
\label{eq:handoff_phase}
\end{equation}
If the product vanishes or no prior beam is active, set $\vartheta_k=0$. The resulting beamformer is
\begin{equation}
\mathbf F_{b,k}=e^{j\vartheta_k}\frac{\mathbf q_k}{\|\mathbf q_k\|_2}
\in\underset{\|\mathbf f\|_2^2\le1}{\arg\max}\,
|\mathbf q_k^{\mathsf H}\mathbf f|^2,
\label{eq:coherent}
\end{equation}
which has unit norm and aligns the two received phases at the predicted handoff position when both responses are nonzero.

\begin{figure}[!t]
\centering
\includegraphics[width=0.86\columnwidth]{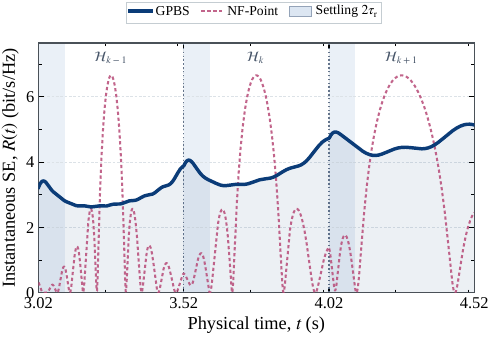}
\caption{Instantaneous SE across consecutive service intervals at $5$~GHz and $M=256$, comparing GPBS against midpoint point focusing ($\gamma_M=20$~dB).}
\label{fig:fig5}
\end{figure}

\begin{samepage}
At $Q$ Gauss--Legendre nodes $t_q$~\cite{GolubWelsch1969}, evaluating and normalizing \eqref{eq:q} takes $O(MQ)$ operations. Explicit assembly of $\mathbf C_k$ and dense eigendecomposition cost $O(M^2Q+M^3)$~\cite{VanTrees2002}. With $\mathbf a_q=\mathbf a(\widetilde{\mathbf z}_k(t_q))$ and normalized weights $w_q$ incorporating the activation kernel, matrix-free power or Lanczos iterations evaluate
\begin{equation}
\mathbf C_k\mathbf v\approx\sum_{q=1}^{Q}w_q\mathbf a_q(\mathbf a_q^{\mathsf H}\mathbf v)
\label{eq:matrix_free_product}
\end{equation}
at $O(MQ)$ per iteration. For $n_k\ge 2K_k$ scalar rows in $\mathbf H_k$, each dense QR fit costs $O(n_kK_k^2)$. Algorithm~\ref{alg:main} performs two fits per decision, which can be accelerated by banded sparse factorization.
\end{samepage}

\par\ifdim\dimexpr\pagegoal-\pagetotal\relax<16\baselineskip\newpage\fi
Proposition~\ref{prop:handoff_gain}, whose proof is in Appendix~\ref{app:handoff_proof}, compares the transition in \eqref{eq:point_gain} with the old and new unit beams $\mathbf f_-=\mathbf f_{b,k-1}$ and $\mathbf f_+=\mathbf F_{b,k}$ at the same user position, with steering vector $\mathbf a=\mathbf a(\mathbf z(t))$.

\begin{proposition}[Coherent Gain During a Beam Handoff]
\label{prop:handoff_gain}
If distinct unit beams produce the same nonzero complex field at the user, every nontrivial normalized blend yields higher gain than either beam alone. In general, let $u_\pm=\mathbf a^{\mathsf H}\mathbf f_\pm$ be the two fields and $g_\pm=|u_\pm|^2$ their gains. For $0<\alpha=\alpha_k(t)<1$ and a nonzero blend in \eqref{eq:point_gain}, its gain $g_{\rm tr}(t)=|\mathbf a^{\mathsf H}\tilde{\mathbf f}_k(t)|^2$ satisfies
\begin{align}
&g_{\rm tr}(t)>\max\{g_-,g_+\}
\quad\Longleftrightarrow\quad \mathcal C_{\rm H}(t).
\label{eq:handoff_gain}\\
\intertext{Let $G=\max\{g_-,g_+\}$ and let $\bar g_\alpha=(1-\alpha)g_-+\alpha g_+$ be the weighted mean of the individual gains. The condition balances normalization against field mismatch and $G-\bar g_\alpha$.}
&\mathcal C_{\rm H}(t)\colon\quad
\alpha(1-\alpha)\bigl(G\|\mathbf f_+-\mathbf f_-\|_2^2
-|u_+-u_-|^2\bigr)>G-\bar g_\alpha.\notag
\end{align}
\end{proposition}

\par\ifdim\dimexpr\pagegoal-\pagetotal\relax<6\baselineskip\newpage\fi
The tail in \eqref{eq:q} gives adjacent beams overlapping design intervals, and \eqref{eq:coherent} aligns their fields at the predicted handoff position. At $t_a$ only the old beam is active for $\tau_{\rm r}>0$. Once mixing begins, \eqref{eq:handoff_gain} tests for a gain along the user's actual path. Figure~\ref{fig:fig5} shows the instantaneous SE across three intervals. Within each beam, motion changes the spherical phase difference $\Delta\varphi_m(t)=2\pi(r_m(t)-r_1(t))/\lambda$. At the kernel mean time $\bar t_k=\int_{\widetilde{\mathcal H}_k} t w_k(t)\,dt$, its rate is
\begin{equation}
\omega_{m,k}=\dot{\Delta\varphi}_m(\bar t_k)
=\frac{2\pi}{\lambda}\bigl(\mathbf u_m-\mathbf u_1\bigr)^{\mathsf T}
\dot{\widetilde{\mathbf z}}_k(\bar t_k),
\label{eq:motion_phase_rate}
\end{equation}
where $\mathbf u_m$ points from element $m$ to the predicted user. Appendix~\ref{app:handoff_proof} relates this rate to the relative aperture weight in \eqref{eq:motion_aperture} and bounds the error due to phase curvature. A boxcar hold under a uniform far-field direction-cosine sweep recovers the zero-offset beam of~\cite{Chen2024Adaptive}.

\section{Simulation Results}
\label{sec:protocol}
\label{sec:results}
We evaluate the communication benefit of turning heterogeneous observations into a motion forecast and one fixed beam for the future service interval. The experiments first compare complete systems, then isolate the contributions of inference and beam shaping, and finally examine their sensitivity to service timing, array size, and motion geometry.

\begin{table}[!t]
\renewcommand{\arraystretch}{1.2}
\caption{Main Simulation Parameters}
\label{tab:protocol}
\centering
\footnotesize
\setlength{\tabcolsep}{4pt}
\begin{tabularx}{\columnwidth}{@{}>{\raggedright\arraybackslash}p{0.46\columnwidth}>{\raggedright\arraybackslash}X@{}}
\toprule
\textbf{Parameter} & \textbf{Setting}\\
\midrule
\multicolumn{2}{@{}l@{}}{\textbf{Link and beam timing}}\\
Carrier, $M$, $d$, $D$ & $(5\,\mathrm{GHz},256,0.03\,\mathrm{m},7.65\,\mathrm{m})$\\
Delay/hold/transition & $(0.02,0.5,0.045)$~s\\
Episode, warm-up time & $[0,6]$~s, $3.0$~s $-T_{\rm h}$\\
SE decision times & $t_k\in\{3.0,3.5,\ldots,5.5\}$~s\\
\addlinespace[2pt]
\multicolumn{2}{@{}l@{}}{\textbf{Communication and sensing observations}}\\
Pilots (rate, SNR per element) & $(40\,\mathrm{Hz},\,10\,\mathrm{dB})$\\
Position (rate, variance, outage) & $(1\,\mathrm{Hz},\,0.09\,\mathrm{m^2},\,[3,6)\,\mathrm{s})$\\
Displ./mean pos. (per./win., var.) & $(2.25,0.04)/(1.5,0.09)\;(\mathrm{s},\mathrm{m^2})$\\
Interval-report arrival lag & $d_j-b_j=0.35$~s\\
\addlinespace[2pt]
\multicolumn{2}{@{}l@{}}{\textbf{Entity inference and prediction}}\\
Spline, smoothness, forecast & $(p,m,N)=(3,3,2)$\\
Knot spacing, smoothness weight & $(1\,\mathrm{s},\,10)$\\
\addlinespace[2pt]
\multicolumn{2}{@{}l@{}}{\textbf{Common Gaussian prior} (diagonal covariance, squared SI units)}\\
Position, velocity means & $(-15,25)^{\mathsf T}$~m, $(4,0)^{\mathsf T}$~m/s\\
Acceleration and jerk means & $\mathbf 0$\\
Per-axis var. (pos./vel./acc./jerk) & $(10^4,400,100,100)$\\
\addlinespace[2pt]
\multicolumn{2}{@{}l@{}}{\textbf{Primary motion}}\\
Midpoint ($t_c=4.5$~s), $\omega_0$ & $(0,25)^{\mathsf T}$~m, $1.25$~rad/s\\
Translation std. dev. (per axis) & $\sigma_\delta=0.08$~m\\
\addlinespace[2pt]
\multicolumn{2}{@{}l@{}}{\textbf{Benchmark protocol}}\\
Nominal speed, downlink SNR & $v=5$~m/s, $\gamma=\gamma_{256}=20$~dB\\
Paired seeds, $Q$, samples/$0.5$~s & $(60,192,640)$\\
Multipath ref. range $\rho_0$, drift std. & $(25\,\mathrm{m},0.6\,\mathrm{m/s})$\\
Scatterer initial region & $[-24,24]\times[10,38]$~m\\
\bottomrule
\end{tabularx}
\end{table}

\begin{figure}[t]
\centering
\includegraphics[width=0.96\columnwidth]{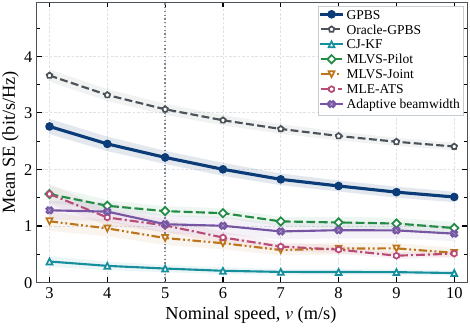}
\caption{Mean SE versus nominal speed. Oracle uses the true future path.}
\label{fig:fig6}
\end{figure}

\subsection{Simulation Setup and Evaluation Protocol}
\label{sec:initialization}
\label{sec:comparators}
\label{sec:supplementary_protocol}
Default settings are listed in Table~\ref{tab:protocol} and varied individually across sweeps. In the primary scenario, the user moves along $\mathbf z(t)=\mathbf z_0+[\Delta x_v(t),0]^{\mathsf T}+\boldsymbol\delta$ through broadside midpoint $\mathbf z_0$ with lateral displacement
\begin{equation}
\Delta x_v(t)=v(t-t_c)+\frac{3v}{4\omega_0}\bigl[\sin(\omega_0(t-t_c)+\phi)-\sin\phi\bigr],
\label{eq:reference_displacement}
\end{equation}
where $\phi\sim\mathcal U(0,2\pi)$ and $\boldsymbol\delta\sim\mathcal N(\mathbf0,\sigma_\delta^2\mathbf I_2)$ are drawn once per seed. Figure~\ref{fig:cdf_2d} evaluates general 2D random trajectories. Both the primary and 2D motion models are specified analytically, independently of the spline basis used for inference.

Each seed provides common motion, channel, and randomized-clock pilot and sensing streams for all methods, which use their respective observation subsets. Each interval-report stream has contiguous, nonoverlapping supports, and the true pilot gains have unit magnitude and are unknown to the estimators. All causal controllers use a common broad Gaussian prior at $t_0$ (or its state marginal), independent of the true motion and speed, with a causal warm-up initializing the previous beam.

We group the baselines in the main figures by observation inputs and beam design to assess fusion, motion inference, and interval coverage. Additional controls are introduced with their corresponding studies.
\begin{figure*}[!t]
\centering
\begin{minipage}[t]{\columnwidth}
\centering
\includegraphics[width=0.96\linewidth]{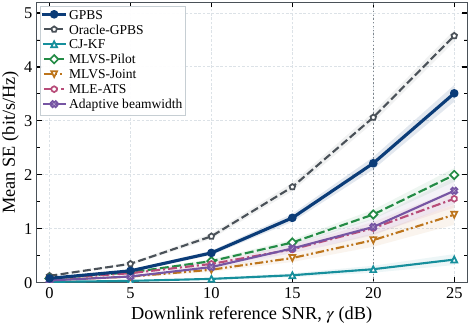}
\caption{Mean SE versus downlink reference SNR.}
\label{fig:fig7}
\end{minipage}\hfill
\begin{minipage}[t]{\columnwidth}
\centering
\includegraphics[width=0.96\linewidth]{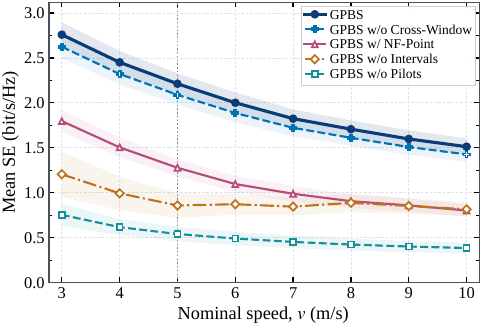}
\caption{Observation and cross-window beam-shaping ablations.}
\label{fig:fig8}
\end{minipage}

\vspace{4pt}
\includegraphics[width=\textwidth]{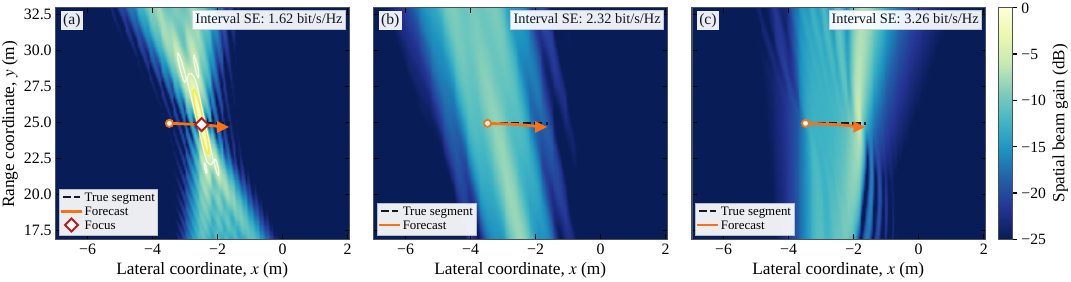}
\caption{Target-beam gains for one entity estimate. (a) Midpoint NF-Point. (b) Far-field Shared-ABW. (c) GPBS. SE includes beam transitions.}
\label{fig:fig9}
\end{figure*}
\par
\par\ifdim\dimexpr\pagegoal-\pagetotal\relax<6\baselineskip\newpage\fi
\noindent\emph{1) Sensing only:} Sparse geometric reports enter through their original temporal operators.\par\nopagebreak[4]
\begin{itemize}
\csname @beginparpenalty\endcsname=10000\relax
\item \textbf{Constant-jerk Kalman filter (\mbox{CJ-KF})}~\cite{Sarkka2023}: Uses full-history smoothing with position and interval reports.
\end{itemize}
\noindent\emph{2) Pilot only:} Dense uplink pilots replace the original echoes, with unknown complex gains eliminated.\par\nopagebreak[4]
\begin{itemize}
\csname @beginparpenalty\endcsname=10000\relax
\item \textbf{Pilot-only maximum-likelihood velocity sensing (\mbox{MLVS-Pilot})}~\cite{Wang2025Velocity}: Fits a local constant-velocity model to the raw pilot likelihoods.
\end{itemize}
\begingroup\interlinepenalty=10000
\noindent\emph{3) Joint communication and sensing:} Pilots are combined with point and interval sensing.\par\nopagebreak[4]
\endgroup
\begin{itemize}
\csname @beginparpenalty\endcsname=10000\relax
\begingroup\interlinepenalty=10000
\item \textbf{Joint maximum-likelihood velocity sensing (\mbox{MLVS-Joint})}~\cite{Wang2025Velocity}: Adds sensing factors to MLVS-Pilot, retaining its constant-velocity model.\par
\endgroup
\item \textbf{Maximum likelihood estimation-based adaptive-parameter Thompson sampling (\mbox{MLE-ATS})}~\cite{Liu2026Thompson}: This adaptation selects a local polar polynomial model and reports a weighted SE score from the maximum-likelihood and posterior-sampled beams, without adaptive feedback scheduling.
\end{itemize}
\noindent\emph{4) Beam-design references:}\par\nopagebreak[4]
\begin{itemize}
\csname @beginparpenalty\endcsname=10000\relax
\item \textbf{Sensing-assisted adaptive beamwidth precoding (Adaptive beamwidth)}~\cite{Chen2024Adaptive}: Using the terminal position and velocity from MLVS-Joint, our implementation directly optimizes angular coverage in the original far-field beam family in place of codebook lookup, retaining the interval rate objective and low-rate penalty.
\item \textbf{Oracle-aided geometrically parameterized beamshaping (\mbox{Oracle-GPBS})}: This noncausal reference integrates spherical responses along the true future path using GPBS's cross-window weighting, handoff alignment, and timing to assess forecast errors.
\end{itemize}

By default, the causal baselines use NF-Point~\cite{Li2024BeamFocusing} at the predicted service midpoint; MLE-ATS applies this rule to its maximum-likelihood and sampled predictions, while Adaptive beamwidth uses the angular-coverage precoder. The primary metric is interval-average SE in \eqref{eq:se} over the complete service interval, weighted by service duration within each seed and averaged across independent seeds. Method comparisons use paired seeds with pointwise $95\%$ Student-$t$ confidence intervals in Figs.~\ref{fig:fig6}--\ref{fig:fig8}, \ref{fig:fig10}--\ref{fig:fig11}, and \ref{fig:delay_benchmark}--\ref{fig:rician_benchmark}, while Section~\ref{sec:supp_source_ablation} adds component studies.

\subsection{Overall Communication Performance}
\label{sec:supp_mc200}
\label{sec:system_performance}
We assess end-to-end performance through two sweeps under a common observation protocol. Figure~\ref{fig:fig6} varies nominal speed at the default SNR to test prediction and coverage. Figure~\ref{fig:fig7} varies downlink SNR at the default speed with fixed pilot quality.

In Fig.~\ref{fig:fig6}, GPBS exceeds the strongest causal baseline by $75\%$ at the default speed, with positive paired pointwise $95\%$ intervals across the sweep. Oracle-GPBS leads GPBS by $0.85$~bit/s/Hz at the same speed, reflecting the effect of forecast errors on beam synthesis and handoffs. As speed increases, both curves decline as the within-hold path grows and differential phase evolution in \eqref{eq:motion_phase_rate} can accelerate, requiring wider near-field coverage. The Oracle trend reflects a power--coverage trade-off even with the true trajectory. Proposition~\ref{prop:fusion_forecast} characterizes the complementary forecast information from pilots and interval reports. MLVS-Joint falls below MLVS-Pilot, consistent with possible interval-report mismatch under its constant-velocity model. Adaptive beamwidth exceeds shared-state MLVS-Joint, consistent with broader angular coverage.

\begin{figure*}[!t]
\centering
\includegraphics[width=\textwidth]{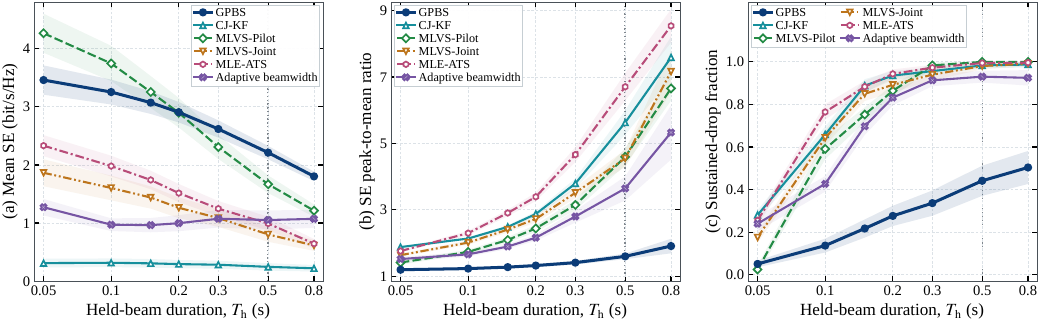}
\caption{Performance versus held-beam duration. (a) Mean SE. (b) SE peak-to-mean ratio. (c) Sustained-drop fraction.}
\label{fig:fig10}
\end{figure*}

In Fig.~\ref{fig:fig7}, Oracle-GPBS, GPBS, and MLVS-Pilot retain their beam sequences as downlink SNR varies. The widening SE gap between Oracle-GPBS and GPBS reflects the increasing rate cost of the same forecast mismatch. At $0$~dB, the paired pointwise $95\%$ interval for the GPBS--MLVS-Pilot difference includes zero, consistent with the near-linear rate dependence on average gain at low effective SNR. At higher SNR, rate concavity penalizes gain fluctuations for a given mean gain, making sustained coverage valuable. GPBS shapes its beam along the predicted path and aligns handoff phases. Proposition~\ref{prop:handoff_gain} specifies when the normalized handoff exceeds both individual-beam gains at the user, and Fig.~\ref{fig:fig10}(b) and (c) examine temporal continuity. Adaptive beamwidth is re-optimized at each SNR.

Together, these results show that forecast accuracy and beam coverage jointly shape communication performance. The gap to Oracle-GPBS highlights the cost of prediction errors, while coverage along the user's path helps sustain gain between updates. The following ablations examine the contributions of observation fusion and beam shaping.

\subsection{Ablation Studies of Fusion and Beam Shaping}
\label{sec:supp_source_ablation}
\label{sec:same_prediction_example}
To examine the sources of the system gains, Fig.~\ref{fig:fig8} varies the observation sources and beam design under common motion, clocks, and service intervals. Figure~\ref{fig:fig9} illustrates spatial coverage for one selected entity estimate.

In Fig.~\ref{fig:fig8}, removing either pilots or interval reports lowers mean SE across the speed sweep, with positive paired \mbox{pointwise} $95\%$ intervals for both losses. The fits retain the common prior, available position reports, and beam-synthesis rule. Pilots provide array-based spatial information, while interval reports constrain historical motion on their native supports. Proposition~\ref{prop:fusion_forecast} characterizes when their forecast information is complementary. With the full entity forecast fixed, midpoint NF-Point also lowers SE relative to GPBS, including the respective beam handoffs. A boxcar target without cross-window weighting or handoff phase alignment lowers SE by $0.12$~bit/s/Hz at the default speed under the same physical transition. This comparison combines both design changes. Proposition~\ref{prop:handoff_gain} gives the conditional criterion for instantaneous handoff gain.

In Fig.~\ref{fig:fig9}, GPBS and NF-Point use the same quadratic forecast, while shared-input adaptive beamwidth (\textbf{\mbox{Shared-ABW}}) uses its terminal position and velocity in a constant-velocity far-field design. NF-Point concentrates gain near the midpoint, whereas Shared-ABW broadens angular coverage. GPBS integrates spherical responses along the predicted path, producing extended near-field coverage. The maps show target-beam gains. The annotated SE includes the transition from the preceding beam.

\subsection{Sensitivity to Service and Motion Conditions}
\label{sec:operating}
Ablations show gains from fusion in the geometric entity and GPBS. Longer holds increase motion between updates, whereas larger arrays sharpen near-field focusing. We vary hold duration and array size separately to examine these effects.
\label{sec:supp_hold}
\label{sec:supp_array_size}
\label{sec:supp_temporal_measures}
\label{sec:supp_hold_continuity}
Figure~\ref{fig:fig10}(a) illustrates the trade-off between focal gain and coverage. At the shortest holds, point focusing concentrates power near the predicted midpoint as the user travels only a short distance, and MLVS-Pilot achieves higher mean SE. Longer holds extend the user's path and favor broader coverage. GPBS exceeds MLVS-Pilot for the tested $T_{\rm h}\geq0.30$~s with positive paired pointwise $95\%$ intervals.

Mean SE cannot distinguish steady rates from brief peaks. The peak-to-mean ratio $P_i$ measures rate concentration, while the sustained-drop fraction $D_i$ captures prolonged peak-relative drops. For hold $j$ of seed $i$, let $\ell_{ij}$, $\bar R_{ij}$, $R^{\max}_{ij}$, and $\tau_{ij}$ denote its duration, mean SE, peak SE, and longest continuous time below half its peak, respectively. The duration-weighted peak-to-mean ratio and sustained-drop fraction are
\begin{equation}
 P_i = \frac{\sum_j\ell_{ij}R^{\max}_{ij}/\bar R_{ij}}
              {\sum_j\ell_{ij}},
 \label{eq:supp_peak_to_mean}
\end{equation}
\begin{equation}
 D_i = \frac{\sum_j\ell_{ij}\mathbf{1}\{\tau_{ij}\geq20~\mathrm{ms}\}}
              {\sum_j\ell_{ij}}.
 \label{eq:supp_sustained_drop}
\end{equation}

Figure~\ref{fig:fig10}(b) and (c) show the seed averages of $P_i$ and $D_i$, respectively. For MLE-ATS, each is also averaged over the maximum-likelihood and posterior-sampled beams using the SE-score weights. As the hold grows, GPBS's peak-to-mean ratio rises more slowly than those of the baselines, and its sustained-drop fraction is lower except at the shortest hold. Longer forecast horizons and spherical phase evolution in \eqref{eq:motion_phase_rate} can give point-focused beams brief peaks and long off-focus periods. GPBS integrates predicted spherical responses in physical time and aligns handoff phases, trading focal gain for extended coverage, consistent with Fig.~\ref{fig:fig8}. Since $D_i$ is peak-relative and low rates can also be flat, both continuity metrics complement mean SE.

\begin{figure}[t]
\centering
\includegraphics[width=0.96\columnwidth]{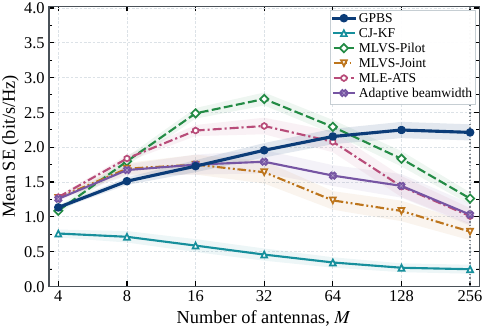}
\caption{Mean SE versus array size with fixed per-element gain.}
\label{fig:fig11}
\vspace{2pt}
\includegraphics[width=0.96\columnwidth]{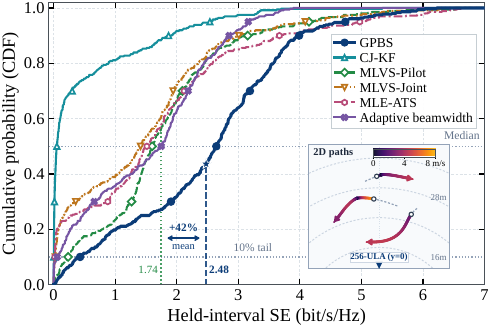}
\caption{Empirical CDF of held-interval SE across $360$ intervals from $60$ paired 2D trajectories. The inset shows representative speed-colored paths.}
\label{fig:cdf_2d}
\end{figure}

In Fig.~\ref{fig:fig11}, larger apertures extend the Rayleigh distance and strengthen spherical near-field effects. Array gain initially helps point focusing, but narrower foci shorten high-gain passages, allowing off-focus losses to offset peak gains. Beyond $M=64$, GPBS leads at $M=128$ and widens its lead at $M=256$, while its own mean SE levels off. Its physical-time integration spreads gain over the predicted path, benefiting mobile near-field XL-MIMO. With small arrays, lower gain reduces GPBS's SE as the link approaches the far field. When responses vary little over the active support, their normalized integral approaches a single steering vector, diminishing the coverage benefit.

\label{sec:supp_signed_curvature}
Figure~\ref{fig:cdf_2d} evaluates $60$ shared 2D paths with random reference positions, headings, and nominal speeds in $[3,7]$~m/s. Smooth speed oscillations and occasional near-stops accompany distance-based turns and oscillatory S-bends. \mbox{Compared with} MLVS-Pilot, the strongest baseline, GPBS achieves $0.73$~bit/s/Hz higher mean SE with a positive paired $95\%$ interval and improves the pooled $10\%$ SE percentile. These gains combine inference and physical-time beam synthesis over the changing geometry.

\begin{figure}[t]
\centering
\includegraphics[width=0.96\columnwidth]{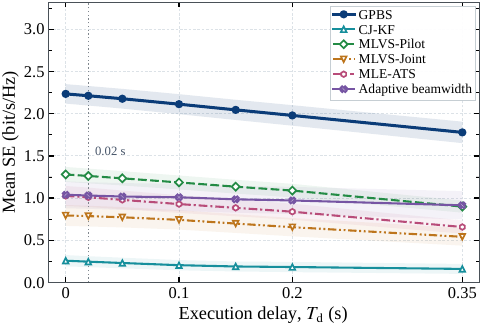}
\caption{Mean SE versus execution delay. The vertical line marks the nominal $0.02$~s delay.}
\label{fig:delay_benchmark}
\end{figure}

\begin{figure}[t]
\centering
\includegraphics[width=0.96\columnwidth]{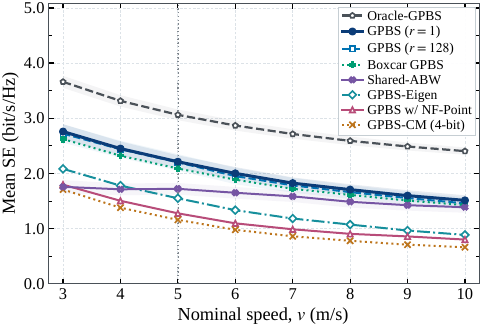}
\caption{Mean SE versus nominal speed for beamforming backends.}
\label{fig:backend_benchmark}
\vspace{2pt}
\includegraphics[width=0.96\columnwidth]{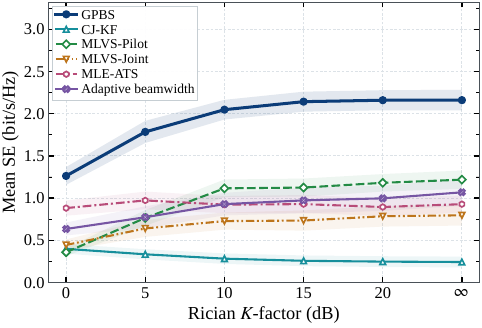}
\caption{Mean SE versus Rician $K$-factor under diffuse near-field multipath and dynamic path loss.}
\label{fig:rician_benchmark}
\end{figure}

\subsection{Execution Delay, Practical Backends, and Multipath}
\label{sec:twc_ext}
\label{sec:scope}
Building on these results, we assess delay, beamforming backends, and multipath over $60$ paired realizations.

Figure~\ref{fig:delay_benchmark} varies $T_{\rm d}$ while retaining each method's observation history and motion estimate. Delay from computation and hardware reconfiguration~\cite{Wang2026MAMovement,Dinis2026FASSwitching,Ting2026LCRIS} shifts the activation window farther ahead. Longer lookahead makes the forecast more sensitive to velocity and acceleration errors. In the near field, even a small position mismatch can move the user away from a narrow focus. GPBS integrates spherical responses along the predicted path to extend coverage. At $T_{\rm d}=0.35$~s, it exceeds the strongest baseline by $0.86$~bit/s/Hz.

Figure~\ref{fig:backend_benchmark} compares beams built from the same geometric estimate, plus Oracle-GPBS with the true path. The dominant-eigenvector beam (\textbf{\mbox{GPBS-Eigen}}) maximizes the predicted kernel-weighted linear gain, whereas GPBS coherently integrates the predicted responses. Both align handoff phases. At the default speed, GPBS achieves $0.66$~bit/s/Hz higher mean SE. The log-rate in \eqref{eq:se} depends on how the realized gain varies over the service interval, including beam transitions. Maximizing predicted linear gain therefore need not maximize actual service SE, especially when the forecast differs from the true path. Changing the reference element alters the relative phase evolution across the aperture and the resulting amplitude taper, although the two reference choices perform similarly here. The unquantized constant-modulus benchmark (\textbf{\mbox{GPBS-CM}}) removes the amplitude taper and reduces SE. Adding $4$-bit phase quantization causes little further loss. In the recorded Python benchmark over $120$ decisions, GPBS synthesis takes $2.31$~ms on average, whereas entity inference and terminal readout take $0.57$~s.

\par\ifdim\dimexpr\pagegoal-\pagetotal\relax<10\baselineskip\newpage\fi
Finally, Fig.~\ref{fig:rician_benchmark} applies diffuse near-field Rician multipath and dynamic path loss to uplink pilots and downlink scoring through
\begin{equation}
\begin{aligned}
\tilde{\mathbf h}(t)=\frac{\rho_0}{\rho(t)}\!\Biggl[&\sqrt{\frac{K}{K+1}}\mathbf a(\mathbf z(t))\\[-2pt]
&\hspace{-1.2em}+\sqrt{\frac{1}{K+1}}\sum_{\ell=1}^{24}\tilde g_\ell e^{-j\frac{2\pi}{\lambda}\Delta r_\ell(t)}\mathbf a(\mathbf s_\ell(t))\Biggr],
\end{aligned}
\label{eq:rician_channel}
\end{equation}
where $\Delta r_\ell(t)=\|\mathbf z(t)-\mathbf s_\ell(t)\|_2-r_1(\mathbf z(t))$, and $\tilde g_\ell$ are complex Gaussian samples normalized to unit total power. The scatterers start uniformly in the region in Table~\ref{tab:protocol} and move at constant velocities drawn independently from a zero-mean isotropic Gaussian. At $K=0$~dB, the LoS and diffuse components have equal nominal power weights and can produce spurious peaks in the LoS-based pilot likelihood. The sensing reports are unaffected by the modeled communication multipath in this simulation and continue to constrain motion when the pilots are distorted. GPBS uses the resulting entity to shape coverage along the predicted path. It exceeds the strongest baseline by $0.38$~bit/s/Hz at $K=0$~dB, and its SE rises with $K$ as the pilot and downlink channels become more LoS-dominant and better aligned with the steering model.

\section{Conclusion}
\label{sec:conclusion}
This paper developed motion-native near-field predictive beam shaping under asynchronous observations, execution delays, and cross-interval transitions. Proposition~\ref{prop:fusion_forecast} identifies when asynchronous pilots and interval reports each contribute information about a future position component given the other source. The refitted terminal derivatives determine the quadratic mean forecast. Cross-window GPBS turns that forecast into a held beam in $O(MQ)$ operations. Proposition~\ref{prop:handoff_gain} gives the condition for a unit-power handoff to exceed both constituent gains at the same user position. Paired simulations show SE and continuity improvements under mobile near-field conditions and characterize sensitivity to speed, SNR, hold duration, array size, delay, and diffuse multipath. GPBS outperforms boxcar GPBS, Shared-ABW, and the dominant eigenbeam in the backend comparison, with a $2.31$~ms mean beam-synthesis time in the recorded Python benchmark. Future work includes beam-dependent acquisition, multi-user scheduling, and compiled sparse entity inference.

\appendices
\section{Pilot-Resolved Directions}
\label{app:pilot_information}
\begin{IEEEproof}[Proof of Proposition~\ref{prop:pilot_information}]
Let $\mathbf a_i=\mathbf a(\bar{\mathbf z}_i)$ and $\boldsymbol\eta_i=(\mathbf y_i-\widehat\alpha_i(\bar{\mathbf z}_i)\mathbf a_i)/\sigma_{\rm p}$. Since $\mathbf a_i^{\mathsf H}\boldsymbol\eta_i=0$, the score and projected information in \eqref{eq:pilot_quadratic} are
\begin{equation}
\begin{aligned}
\mathbf J_i&=-\frac{\widehat\alpha_i(\bar{\mathbf z}_i)}{\sigma_{\rm p}}
(\mathbf I_M-\mathbf a_i\mathbf a_i^{\mathsf H})
\left.\frac{\partial\mathbf a}{\partial\mathbf z^{\mathsf T}}\right|_{\bar{\mathbf z}_i},\\
\mathbf g_i&=2\operatorname{Re}(\mathbf J_i^{\mathsf H}\boldsymbol\eta_i),\quad
\mathbf I_i=2\operatorname{Re}(\mathbf J_i^{\mathsf H}\mathbf J_i).
\end{aligned}
\label{eq:raw_jacobian}
\end{equation}
For real $\mathbf v$, $\mathbf v^{\mathsf T}\mathbf I_i\mathbf v=2\|\mathbf J_i\mathbf v\|_2^2$. Since $\widehat\alpha_i\ne0$, \eqref{eq:pilot_information_null} follows.
\end{IEEEproof}

\par\ifdim\dimexpr\pagegoal-\pagetotal\relax<5\baselineskip\newpage\fi
\noindent\emph{Association and screening.} For initial association, a constant-acceleration sensing smoother ($q=10$) provides the guide $(\mathbf p_{0,i},\mathbf P_{0,i})$ at $t_i$. Its covariance is regularized by adding $0.08^2\mathbf I_2$ and clipping its eigenvalues to $[0.08^2,0.35^2]$~m$^2$. The bounded search reduces
\begin{equation*}
\psi_i(\mathbf z)+\tfrac12\|\mathbf z-\mathbf p_{0,i}\|^2_{\mathbf P_{0,i}^{-1}},
\end{equation*}
where $\psi_i(\mathbf z)=\delta^2(\sqrt{1+\ell_i(\mathbf z)/\delta^2}-1)$ is the pseudo-Huber pilot cost with $\delta=2$. Within $0.20$~m of the guide, a $3\times3$ grid and up to $3$ Gauss--Newton steps are used, with a $0.08$~m step cap and up to $4$ halvings. The guide penalty selects the expansion point and is not added to the pilot factor. Let $c_0=\psi_i(\mathbf p_{0,i})$, $c_1=\psi_i(\bar{\mathbf z}_i)$, $\Delta c_i=\max(c_0-c_1,0)$, and $d_i^2=\|\bar{\mathbf z}_i-\mathbf p_{0,i}\|^2_{\mathbf P_{0,i}^{-1}}$. The association weight is
\begin{equation*}
\omega_i=\Bigl(1-e^{-\Delta c_i/(0.05\max(c_0,1))}\Bigr)(1+d_i^2)^{-1}.
\end{equation*}
Pilot $i$ enters $\mathcal I_k$ if $\Delta c_i\ge 10^{-4}$, $\omega_i\ge 0.03$, $\lambda_{\min}(\mathbf I_i)\ge 10^{-10}$, and $\kappa(\mathbf I_i)\le 10^8$, with $w_i=0.35\,\omega_i$ in \eqref{eq:common_factor}.

\par\ifdim\dimexpr\pagegoal-\pagetotal\relax<10\baselineskip\newpage\fi
\section{Complementary Forecast Information}
\label{app:fusion_proof}
\begin{IEEEproof}[Proof of Proposition~\ref{prop:fusion_forecast}]
For a lookahead $\tau$, unit direction $\mathbf u$, and a historical perturbation $h(t)=\mathbf B_k(t)\mathbf v$, the order-$N$ Taylor readout in \eqref{eq:taylor_forecast} is $\ell_N(h)=\sum_{n=0}^N(\tau^n/n!)\,\mathbf u^{\mathsf T}h^{(n)}(t_k)=\mathbf d_N^{\mathsf T}\mathbf v$, where $\mathbf d_N=\mathbf D_{N,k}^{\mathsf T}\mathbf E_N^{\mathsf T}(\tau)\mathbf u$. For either single-source fit $r\in\{\mathrm P,\mathrm S\}$, let $\mathbf P_r$ be the coefficient covariance and $\bar r$ the other source. Then $\sigma_r^2=\mathbf d_N^{\mathsf T}\mathbf P_r\mathbf d_N$ and $\boldsymbol\kappa_{r,N}(s)=\mathbf B_k(s)\mathbf P_r\mathbf d_N$.

All three fits retain the same point-position reports and the boundary and smoothness costs in \eqref{eq:init_cost}--\eqref{eq:reg_cost}. The regularization is positive definite because zero cost forces $h^{(m)}$ to vanish on each knot span and the initial derivatives to vanish. Continuity through order $m-1$ then gives $h=0$. Write $\mathbf H_{\rm P}$ for the stack of accepted pilot blocks $\sqrt{w_i}\mathbf I_i^{1/2}\mathbf B_k(t_i)$ and $\mathbf H_{\rm S}$ for the stack of arrived interval blocks $(\mathbf R_j^{\rm s})^{-1/2}\mathcal L_j[\mathbf B_k]$ with $a_j<b_j$. We fix the accepted pilot indices, weights, and expansion points for these comparisons. The refinement in \eqref{eq:feedback_seed} replaces each updated factor.

Independent errors give $\mathbf P_{\rm P+S}^{-1}=\mathbf P_r^{-1}+\mathbf H_{\bar r}^{\mathsf T}\mathbf H_{\bar r}$. The matrix inversion lemma yields
\begin{equation*}
\begin{aligned}
\sigma_r^2-\sigma_{\rm P+S}^2
 &=\mathbf b^{\mathsf T}
 (\mathbf I+\mathbf H_{\bar r}\mathbf P_r\mathbf H_{\bar r}^{\mathsf T})^{-1}\mathbf b,\\
\mathbf b&=\mathbf H_{\bar r}\mathbf P_r\mathbf d_N .
\end{aligned}
\end{equation*}
The inverse is positive definite, so $\sigma_{\rm P+S}^2\le\min\{\sigma_{\rm P}^2,\sigma_{\rm S}^2\}$ always holds. The variance decreases strictly from fit $r$ exactly when at least one block of $\mathbf b$ is nonzero. Starting from the interval fit, the blocks are $\sqrt{w_i}\mathbf I_i^{1/2}\boldsymbol\kappa_{\rm S,N}(t_i)$. Starting from the pilot fit, they are $(\mathbf R_j^{\rm s})^{-1/2}\mathcal L_j[\boldsymbol\kappa_{\rm P,N}]$. Since $w_i>0$ and $\mathbf R_j^{\rm s}\succ0$, the two strict decreases hold exactly under $\mathcal C_{\rm P\mid S}$ and $\mathcal C_{\rm S\mid P}$, respectively. Their conjunction gives \eqref{eq:fusion_forecast_precision}. Equation~\eqref{eq:operators} evaluates the sensing test at the displacement endpoints or over the averaging interval, even when the report arrives later.

For affine motion, a position at $a$ and the displacement $\Delta\mathbf z$ on $[a,b]$ determine the forecast $\mathbf z(a)+(t-a)\Delta\mathbf z/(b-a)$.
\end{IEEEproof}

\par\ifdim\dimexpr\pagegoal-\pagetotal\relax<10\baselineskip\newpage\fi
\section{Coherent Handoff and Aperture Response}
\label{app:handoff_proof}
\begin{IEEEproof}[Proof of Proposition~\ref{prop:handoff_gain}]
Let $s=\alpha(1-\alpha)>0$ and let $D>0$ be the square of the denominator in \eqref{eq:point_gain}. Expanding gives $D=1-s\|\mathbf f_+-\mathbf f_-\|_2^2$ and $Dg_{\rm tr}(t)=\bar g_\alpha-s|u_+-u_-|^2$. Thus $g_{\rm tr}(t)>G$ exactly when $\mathcal C_{\rm H}(t)$ holds. If $u_-=u_+\ne0$ for distinct beams, then $Dg_{\rm tr}(t)=G>0$ and $D<1$, giving $g_{\rm tr}(t)>G$.
\end{IEEEproof}

\smallskip
\noindent\emph{Aperture weights of the constituent beams.}
Write $\widehat w_k(\omega)=\int_{\widetilde{\mathcal H}_k} w_k(t)e^{-j\omega(t-\bar t_k)}dt$. Since $[\mathbf a]_1=1/\sqrt M$, normalization and handoff rotation in \eqref{eq:coherent} cancel in the ratio
\begin{equation}
\begin{aligned}
\frac{[\mathbf F_{b,k}]_m}{[\mathbf F_{b,k}]_1}
 &=\int_{\widetilde{\mathcal H}_k} w_k(t)e^{-j\Delta\varphi_m(t)}\,dt\\
 &=e^{-j\Delta\varphi_m(\bar t_k)}\widehat w_k(\omega_{m,k})+\varepsilon_{m,k}.
\end{aligned}
\label{eq:motion_aperture}
\end{equation}
The ratio has unit modulus exactly when the phasor is constant $w_k$-almost everywhere. For $|\ddot{\Delta\varphi}_m|\le L_{m,k}$, Taylor's theorem at $\bar t_k$ and $|e^{-jx}-e^{-jy}|\le |x-y|$ give $|\varepsilon_{m,k}|\le L_{m,k}\operatorname{Var}_{w_k}(t)/2$. The error vanishes for affine phase.

For $r_m>0$, $\dot r_m=\mathbf u_m^{\mathsf T}\dot{\widetilde{\mathbf z}}_k$ gives \eqref{eq:motion_phase_rate}. Reference element $r$ gives $\mathbf q_r=\sqrt M\,\mathbf C_k\mathbf e_r$ and replaces $\mathbf u_1$ by $\mathbf u_r$. A boxcar hold gives $\widehat w_k(\omega)=\operatorname{sinc}(\omega T_{\rm h}/(2\pi))$ with $\operatorname{sinc}(v)=\sin(\pi v)/(\pi v)$. For $a_m(u)=e^{-j(m-1)\pi u}/\sqrt M$ and a uniform far-field sweep on $[u_0-\delta,u_0+\delta]$, the ratio is $e^{-j(m-1)\pi u_0}\operatorname{sinc}((m-1)\delta)$.

For the gain comparison in Fig.~\ref{fig:backend_benchmark}, \eqref{eq:coherent_rate_bound} at $\mathbf f=\mathbf F_{b,k}$ and the Rayleigh bound give $\|\mathbf q_k\|_2^2\le\mathbf F_{b,k}^{\mathsf H}\mathbf C_k\mathbf F_{b,k}\le\lambda_{\max}(\mathbf C_k)$. The upper endpoint is attained when $\mathbf q_k$ lies in the leading eigenspace, including the case of a constant steering response.

\IEEEtriggeratref{8}


\begin{thebibliography}{10}
\providecommand{\url}[1]{#1}
\csname url@samestyle\endcsname
\providecommand{\newblock}{\relax}
\providecommand{\bibinfo}[2]{#2}
\providecommand{\BIBentrySTDinterwordspacing}{\spaceskip=0pt\relax}
\providecommand{\BIBentryALTinterwordstretchfactor}{4}
\providecommand{\BIBentryALTinterwordspacing}{\spaceskip=\fontdimen2\font plus
\BIBentryALTinterwordstretchfactor\fontdimen3\font minus \fontdimen4\font\relax}
\providecommand{\BIBforeignlanguage}[2]{{%
\expandafter\ifx\csname l@#1\endcsname\relax
\typeout{** WARNING: IEEEtran.bst: No hyphenation pattern has been}%
\typeout{** loaded for the language `#1'. Using the pattern for}%
\typeout{** the default language instead.}%
\else
\language=\csname l@#1\endcsname
\fi
#2}}
\providecommand{\BIBdecl}{\relax}
\BIBdecl

\bibitem{Wang2024XLMIMO}
Z.~Wang, J.~Zhang, H.~Du, D.~Niyato, S.~Cui, B.~Ai, M.~Debbah, K.~B. Letaief, and H.~V. Poor, ``A tutorial on extremely large-scale {MIMO} for {6G}: Fundamentals, signal processing, and applications,'' \emph{IEEE Commun. Surveys Tuts.}, vol.~26, no.~3, pp. 1560--1605, 2024.

\bibitem{Lu2024Tutorial}
H.~Lu, Y.~Zeng, C.~You, Y.~Han, J.~Zhang, Z.~Wang, Z.~Dong, S.~Jin, C.-X. Wang, T.~Jiang, X.~You, and R.~Zhang, ``A tutorial on near-field {XL-MIMO} communications toward {6G},'' \emph{IEEE Commun. Surveys Tuts.}, vol.~26, no.~4, pp. 2213--2257, 2024.

\bibitem{Cui2023NearField}
M.~Cui, Z.~Wu, Y.~Lu, X.~Wei, and L.~Dai, ``Near-field {MIMO} communications for {6G}: Fundamentals, challenges, potentials, and future directions,'' \emph{IEEE Commun. Mag.}, vol.~61, no.~1, pp. 40--46, Jan. 2023.

\bibitem{Nazar2026Enwar3}
A.~M. Nazar, A.~Celik, A.~Abdallah, M.~Y. Selim, D.~Qiao, and A.~M. Eltawil, ``{Enwar 3.0}: An agentic multi-modal {LLM} orchestrator for situation-aware beamforming, blockage prediction, and handover management,'' \emph{IEEE Trans. Mobile Comput.}, vol.~25, no.~11, pp. 19\,158--19\,175, Nov. 2026, {doi}: \href{https://doi.org/10.1109/TMC.2026.3709048}{\nolinkurl{10.1109/TMC.2026.3709048}}.

\bibitem{Wang2026MAMovement}
H.~Wang, Q.~Wu, Y.~Gao, W.~Chen, W.~Mei, G.~Hu, and L.~Xu, ``Throughput maximization for movable antenna systems with movement delay consideration,'' \emph{IEEE Trans. Wireless Commun.}, vol.~25, pp. 883--899, 2026.

\bibitem{Wang2024LiquidCrystal}
P.-Y. Wang, B.~Sievert, A.~Rennings, and D.~Erni, ``A liquid-crystal-based dynamic metasurface antenna for electronical beam-steering at 105 {GHz},'' \emph{IEEE Trans. Antennas Propag.}, vol.~72, no.~6, pp. 5313--5318, Jun. 2024.

\bibitem{Dinis2026FASSwitching}
D.~Dinis and R.~Wichman, ``Spatio-temporal port selection in fluid antennas under switching delays,'' \emph{IEEE Commun. Lett.}, vol.~30, pp. 992--996, 2026.

\bibitem{Feng2025BeamManagement}
Y.~Feng, P.~Dinh, M.~Ghoshal, E.~Baena, H.~Bouchebbah, and D.~Koutsonikolas, ``Vivisecting beam management in operational {5G} {mmWave} networks,'' \emph{Proc. ACM Netw.}, vol.~3, no. CoNEXT2, pp. 12:1--12:26, Jun. 2025.

\bibitem{Hernandez2024GNSSLatency}
J.~Hern{\'a}ndez~Olcina, A.~B. Anquela~Juli{\'a}n, and {\'A}.~E. Mart{\'i}n~Furones, ``Navigating latency hurdles: An in-depth examination of a cloud-powered {GNSS} real-time positioning application on mobile devices,'' \emph{Sci. Rep.}, vol.~14, p. 14668, 2024.

\bibitem{Urbinati2026RISLatency}
L.~Urbinati, N.~Decarli, F.~Guidi, A.~Guerra, B.~M. Masini, and A.~Zanella, ``On-device deep learning for {RIS} beamforming: Design and latency-rate trade-offs,'' \emph{IEEE Open J. Commun. Soc.}, vol.~7, pp. 4605--4622, 2026.

\bibitem{Biswas2026VIBE}
A.~Biswas, A.~Pramanik, E.~Ekici, and M.~C. Vuran, ``Look once, beam twice: Camera-primed real-time double-directional {mmWave} beam management for vehicular connectivity,'' in \emph{Proc. IEEE SECON}, 2026, pp. 533--542.

\bibitem{Ting2026LCRIS}
T.-L. Ting, Y.-C. Chang, Y.-C. Hung, C.-Y. Hsiao, C.-K. Wen, and T.-H. Lin, ``Liquid-crystal-based reconfigurable intelligent surface for 7 {GHz} mobile communication,'' \emph{Adv. Mater. Technol.}, vol.~11, no.~22, Nov. 2026, {Art.} no.~e71295.

\bibitem{Zhang2026NearFieldDynamics}
X.~Zhang, H.~Zhang, and Y.~C. Eldar, ``Near-field time-varying channel: Analysis and tracking,'' \emph{IEEE Trans. Wireless Commun.}, vol.~25, pp. 9280--9294, 2026.

\bibitem{Wang2025Velocity}
Z.~Wang, X.~Mu, and Y.~Liu, ``Near-field velocity sensing and predictive beamforming,'' \emph{IEEE Trans. Veh. Technol.}, vol.~74, no.~1, pp. 1806--1810, Jan. 2025.

\bibitem{Jiang2024NISE}
H.~Jiang, Z.~Wang, and Y.~Liu, ``Near-field sensing enabled predictive beamforming: From estimation to tracking,'' \emph{IEEE Trans. Commun.}, 2026, early access, doi: \href{https://doi.org/10.1109/TCOMM.2026.3738484}{\nolinkurl{10.1109/TCOMM.2026.3738484}}.

\bibitem{Liu2026Thompson}
J.~Liu, Z.~Wang, S.~Tsai, and R.~Zhang, ``Adaptive payload-aided near-field beam tracking via {Thompson} sampling,'' \emph{arXiv preprint \href{https://arxiv.org/abs/2608.21770}{arXiv:2608.21770}}, 2026.

\bibitem{Chen2024Adaptive}
W.~Chen, Y.~Fan, C.~Yang, and M.~Peng, ``Sensing-assisted beam tracking with real-time beamwidth adaptation for {THz} communications,'' \emph{IEEE Trans. Veh. Technol.}, vol.~75, no.~7, pp. 13\,551--13\,559, Jul. 2026, {doi}: \href{https://doi.org/10.1109/TVT.2026.3658550}{\nolinkurl{10.1109/TVT.2026.3658550}}.

\bibitem{Ye2025TABS}
S.~Ye, Y.~Gao, M.~Xiao, P.~Wang, M.~Poulakis, and U.~Imberg, ``Trajectory-adaptive beam shaping: Towards beam-management-free near-field communications,'' in \emph{Proc. IEEE Int. Conf. Commun. (ICC)}, Glasgow, U.K., May 2026, pp. 1--6, {doi}: \href{https://doi.org/10.1109/ICC59461.2026.11587600}{\nolinkurl{10.1109/ICC59461.2026.11587600}}.

\bibitem{Fan2026GeometricTracking}
Y.~Fan, C.~Yang, and M.~Peng, ``From discrete snapshots to geometric entity: Motion-native channel tracking for {THz} near-field {XL-MIMO},'' \emph{arXiv preprint \href{https://arxiv.org/abs/2611.10587}{arXiv:2611.10587}}, 2026.

\bibitem{Barfoot2014SparseGP}
T.~D. Barfoot, C.~H. Tong, and S.~S{\"a}rkk{\"a}, ``Batch continuous-time trajectory estimation as exactly sparse {G}aussian process regression,'' in \emph{Proceedings of Robotics: Science and Systems}, 2014.

\bibitem{Johnson2024Continuous}
J.~C. Johnson, J.~G. Mangelson, T.~D. Barfoot, and R.~W. Beard, ``Continuous-time trajectory estimation: A comparative study between {G}aussian process and spline-based approaches,'' \emph{arXiv preprint \href{https://arxiv.org/abs/2402.00399}{arXiv:2402.00399}}, 2024.

\bibitem{deBoor2001}
C.~de~Boor, \emph{A Practical Guide to Splines}.\hskip 1em plus 0.5em minus 0.4em\relax New York, NY, USA: Springer, 2001.

\bibitem{GolubPereyra1973}
G.~H. Golub and V.~Pereyra, ``The differentiation of pseudo-inverses and nonlinear least squares problems whose variables separate,'' \emph{SIAM Journal on Numerical Analysis}, vol.~10, no.~2, pp. 413--432, Apr. 1973.

\bibitem{GolubWelsch1969}
G.~H. Golub and J.~H. Welsch, ``Calculation of {G}auss quadrature rules,'' \emph{Mathematics of Computation}, vol.~23, no. 106, pp. 221--230, Apr. 1969.

\bibitem{VanTrees2002}
H.~L. Van~Trees, \emph{Optimum Array Processing}.\hskip 1em plus 0.5em minus 0.4em\relax New York, NY, USA: Wiley, 2002.

\bibitem{Sarkka2023}
S.~S{\"a}rkk{\"a} and L.~Svensson, \emph{Bayesian Filtering and Smoothing}, 2nd~ed.\hskip 1em plus 0.5em minus 0.4em\relax Cambridge, U.K.: Cambridge University Press, 2023.

\bibitem{Li2024BeamFocusing}
X.~Li, Z.~Dong, Y.~Zeng, S.~Jin, and R.~Zhang, ``Multi-user modular {XL-MIMO} communications: Near-field beam focusing pattern and user grouping,'' \emph{IEEE Trans. Wireless Commun.}, vol.~23, no.~10, pp. 13\,766--13\,781, Oct. 2024.

\end{thebibliography}
\end{document}